\documentclass[journal]{IEEEtran}
\usepackage{amsmath}
\usepackage{amssymb}
\usepackage{graphicx}
\usepackage{hyperref}
\usepackage{color}
\usepackage{soul}
\hypersetup{bookmarks,
           colorlinks,
            citecolor=red,
            filecolor=blue,
            urlcolor=blue}

\ifCLASSINFOpdf
\else
\fi

\begin{document}

\title{Frequency Precision of Oscillators Based on High-Q Resonators}

\author{Eyal Kenig and M.~C.~Cross
\thanks{The authors are with the Department
of Physics, California Institute of Technology, Pasadena,
CA, 91125 USA (e-mail: mcc@caltech.edu).}}

\maketitle

\begin{abstract}
We present a method for analyzing the phase noise of oscillators based on feedback driven high quality factor resonators. Our approach is to derive the phase drift of the oscillator by projecting the stochastic oscillator dynamics onto a slow time scale corresponding physically to the long relaxation time of the resonator. We derive general expressions for the phase drift generated by noise sources in the electronic feedback loop of the oscillator. These are mixed with the signal through the nonlinear amplifier, which makes them {cyclostationary}. We also consider noise sources acting directly on the resonator. The expressions allow us to investigate reducing the oscillator phase noise thereby improving the frequency precision using resonator nonlinearity by tuning to special operating points. We illustrate the approach giving explicit results for a phenomenological amplifier model. We also propose a scheme for measuring the slow feedback noise generated by the feedback components in an open-loop driven configuration in experiment or using circuit simulators, which enables the calculation of the closed-loop oscillator phase noise in practical systems.
\end{abstract}

\IEEEpeerreviewmaketitle

\section{Introduction}

\IEEEPARstart{S}{elf}-sustained oscillators have a major technological significance. Such devices, generating a
periodic signal at an inherent frequency, are often developed to
serve as highly accurate time or frequency references \cite{vig}.

In this paper, we present a systematic formalism for calculating the frequency precision of oscillators comprised of a high quality factor (Q) resonator driven by a sustaining electronic feedback loop. This type of architecture is common in time and frequency references, such as quartz crystal or MEMS based systems. The high-Q resonator provides the basic frequency determining element; the electronic feedback system injects the energy needed to sustain the motion without perturbing the resonator frequency too much. The intuition is that increasing the Q of the resonator improves the frequency stability, and this is confirmed by the Leeson analysis \cite{Leeson66}, which provides the standard expression for quantifying the performance. The sustained motion forms a limit cycle in the phase space of dynamical variables of the system; a limit cycle in a deterministic system is purely periodic, and would have perfect frequency precision. Deviations from this simple description are due to \emph{noise} in the system, which may come from thermal, electronic, vibrational or other sources. Thus the analysis of the frequency precision of oscillators requires the calculation of the effect of stochastic terms in the dynamics.

An important concept in describing an oscillator is the \emph{phase} variable $\Phi$. This can be thought of as the angle defining the position of the phase space point around the limit cycle. By a suitable (nonlinear) transformation of variables, the limit cycle can be rendered circular, with the phase advancing uniformly in time in the deterministic system. The frequency of the oscillator is then given by the constant rate of advancement of the phase $\omega=\dot\Phi$ with the dot denoting the time derivative $d/dt$. Since oscillators are sustained by a feedback mechanism, and not by an external clock, they possess a phase invariance property which makes the phase sensitive to stochastic perturbations. The stochastic phase dynamics broaden the peaks in the power spectrum of the oscillator output representing the periodic motion of the limit cycle, and degrade its performance.

General schemes have been developed to calculate the stochastic phase dynamics and resulting precision degradation of oscillators \cite{DemirMehrotra00,Demir02,DemirRoychowdhury03,SuvakDemir11}. However, these require complex numerical implementation. The numerical calculations are made more difficult in the case of high-Q resonators due to the disparate time scales in the system: the relaxation rate towards the limit cycle, which will typically be of order $\omega/Q$ with $Q$ the resonator quality factor (perhaps modified by the loading of the feedback system), leading to a relaxation time of order $Q$ times the period of the oscillator. We are particularly interested in situations where the resonator is driven into its nonlinear regime, where the frequency becomes dependent on the amplitude of oscillation. This regime becomes increasingly important as devices are made smaller, so that the amplitude of motion must be increased to be readily detected, and has been suggested to be important in various noise suppression techniques. Driving the resonator into the regime of nonlinear dynamics further adds to the difficulty of numerical solution. Root-finding methods, rather than direct time simulations, have the problem of multiplicity of solutions, and the need to investigate the stability of the different solutions. An additional drawback is that it may be hard to discern the dependence on system parameters, without an exhaustive set of calculations, and little intuition is gained that might help in the design of improved performance.

Our analysis is made by focusing on the complex envelope function describing the oscillatory motion in terms of the slow modulation of oscillations at the linear resonance frequency of the resonator (which we call the \emph{carrier frequency}), a common and widely used method for analyzing weakly nonlinear systems \cite{crossBook,strogatz,LCreview}. Since the high-Q resonator acts as a strong filter, we can calculate the effects of the feedback system by focusing on the output of the amplifier (both deterministic and stochastic) at the carrier frequency. The behavior of the amplifier system will in general not have a strong dependence on frequency (i.e., the behavior will effectively be constant over the band of frequencies of order $\omega/Q$, characteristic of the width of the resonator response), and we can characterize the performance as if the input to the amplifier were periodic, neglecting the slow time dependence of the envelope function. Thus our approach combines two widely used methods: the envelope formalism for describing the interesting weakly nonlinear behavior of driven resonators, with the analysis of amplifier performance for periodic input signals. This approach formalizes the intuitive way of understanding feedback oscillators in terms of resonator behavior driven by feedback characterized by an amplitude, a phase, and some noise. As a result, the dependence on the variety of system parameters becomes evident, and developing ideas to suppress the degradation due to noise becomes easier. Experimental results related to the ideas presented here can be found in Ref.~\cite{Villanueva13}.

Complex envelope function approaches have been used before to discuss noise properties of oscillators, see for example Refs.~\cite{Lax67,greywall,YurkePra}. The novelty of our work is in combining a complex envelope description of the resonator with a full treatment of realistic feedback systems, including the possibility of strong nonlinearity leading to complicated statistics of the feedback noise. An important result is that we show how to reconcile the cyclostationarity of the amplifier noise with a periodic input signal with the fact that the statistics of the oscillator noise must be stationary, since there is no time reference for a free running oscillator.

The outline of the paper is as follows. In the next section we elaborate on the architecture of oscillators we consider and establish the basic ideas and methods of the approach. We then derive the complex envelope noise for various noise sources in the oscillator system, and describe general schemes for eliminating or reducing the oscillator phase noise. In the following section we apply these methods to a phenomenological model of amplifiers represented by a nonlinear gain function, and describe specific methods for improving oscillator performance by choosing optimal operating points. Finally we consider the application of the method to more realistic models of the amplifier, including a way to simulate or experimentally measure the components of the feedback noise relevant to the oscillator performance using an externally driven, open-loop system. Details of the calculations are deferred to appendices.

\section{Oscillators in the envelope formalism}

\subsection{Basic setup}\label{subsec: Basic setup}

\begin{figure}[tbh]
  \includegraphics[width=0.95\columnwidth]{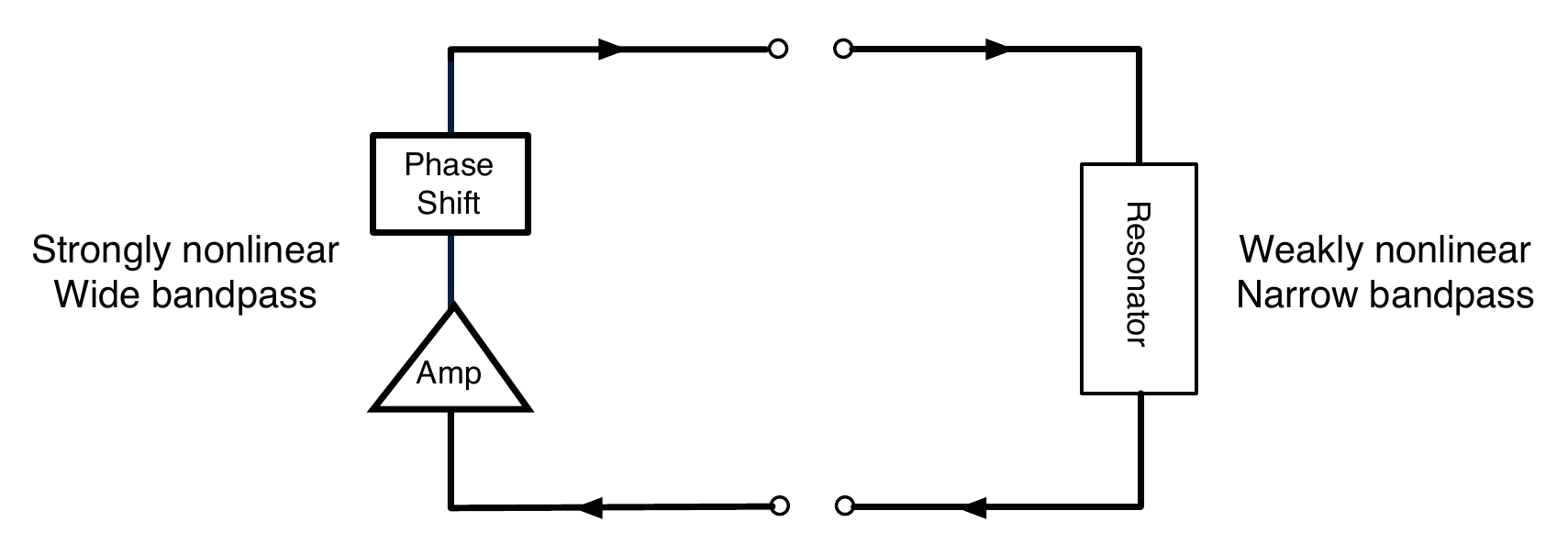}
  \caption{\label{figmcc}  Architecture of a precision oscillator, consisting of a resonator, a sustaining amplifier, and a phase shifter.}
\end{figure}

Figure \ref{figmcc} gives a general schematic of the oscillator architecture we consider. It consists of a resonator driven by a feedback loop containing an amplifier and a phase shifter.

The resonator is described by an equation of motion of the form
\begin{equation}
\ddot q+Q^{-1}\dot q+q+bq^{3}+c q^{2}\dot q=Q^{-1}d(t).\label{resonatorEOM}
\end{equation}
We have scaled time in units of $\omega_{0}^{-1}$, with $\omega_{0}$ the linear resonance frequency, so that the resonance frequency in the scaled units is $1$.  The second term on the left hand side of Eq.~(\ref{resonatorEOM}) is the linear dissipation and introduces the quality factor $Q$. The term $Q^{-1}d(t)$ on the right hand side is the driving force. In the closed loop oscillator the drive will come from the feedback and must balance the dissipation for sustained oscillations; we have included the explicit factor of $Q^{-1}$ in the drive term so that oscillations onset for $d=O(1)$. We include the noise forces by taking $d(t)\to d(t)+\xi(t)$ with $\xi(t)$ a stochastic variable. The nonlinear term $bq^{3}$ acts to shift the resonant frequency as the amplitude of oscillation grows. We have also included a nonlinear correction to the dissipation $c q^{2}\dot q$ which may also be present \cite{dykmanNonlinearDamping, LCreview}. We will phrase much of the discussion in the context of a mechanical resonator, such as a quartz element or a MEMS or NEMS device, for which $q$ is the displacement in a particular mode, but the results apply equally well to electrical or other resonators.

Precision oscillators are typically constructed from resonators with large values of $Q$, and we develop our approximate treatment of the resonator through an expansion in the small parameter $\varepsilon=Q^{-1}$. We treat the dynamics of the resonator by introducing a slowly varying complex envelope $A=ae^{i\Phi}$ with magnitude $a$ and phase $\Phi$ modulating the oscillations at the resonance frequency, writing the output signal of the resonator as
\begin{equation}\label{signal}
    q(t)=\tfrac{1}{2}A(T)e^{it}+\text{c.c.}+O(\varepsilon),
\end{equation}
with $T=\varepsilon t$ a dimensionless slow time scale. The symbol c.c.\ denotes the complex conjugate. From Eq.~(\ref{signal}), the complex amplitude $A(T)$ is obtained from $q(t)$ by averaging over a period
\begin{equation}\label{A average}
A(T) =\frac{1}{ \pi}\int_{\varepsilon^{-1}T- \pi}^{\varepsilon^{-1}T+ \pi}q(t)e^{-it}dt.
\end{equation}
The $O(\varepsilon)$ terms in Eq.~(\ref{signal}) represent higher order terms in the expansion, including, for example, harmonics.

Noise suppression using a nonlinear resonator occurs when the resonator is driven hard enough so that the change in the frequency due to the dependence of the frequency on amplitude is comparable to the line width of the linear resonance spectrum, given by the dissipation in the resonator \cite{greywall,YurkePra,kenig12,nonlinearCrystal}. For a high-Q resonator the line width is much less than the frequency itself, so that the resonator remains \emph{weakly} nonlinear under these conditions, even though the resonator response for a fixed drive level and frequency may show complex behavior such as a multiplicity of solutions \cite{LCreview}. The weak nonlinearity means that the frequency change is small compared to the resonance frequency and harmonic production is small. The weak nonlinearity is introduced into the formalism by supposing the nonlinear coefficient $b$ to be $O(\varepsilon)$, so that for $q=O(1)$ the nonlinear frequency pulling is comparable to the linear resonator line width. The $O(\varepsilon)$ terms in Eq.~(\ref{signal}) include higher harmonics generated by this nonlinearity.

The amplifier in the sustaining feedback loop, on the other hand, may be strongly nonlinear, producing harmonics of its input signal, and up- and down-conversion of noise by mixing with the signal. However
the frequency response of the feedback system will typically be broad compared with the line width of the resonator, i.e. the output of the feedback is approximately constant as a function of frequency over the relevant frequency range. This means that in considering the effect of the feedback loop on the resonator, we may \emph{ignore} the slow time dependence $A(T)$, and effectively study the behavior for a \emph{periodic} input to the amplifier at unit frequency (frequency $\omega_{0}$ in unscaled units). Note that the oscillator will not usually operate at exactly this frequency, but sufficiently close to this so that the behavior of the amplifier will not be significantly different. The output of the feedback system that drives the resonator will then be periodic at the same frequency, but now including harmonics, together with noise which will appear \emph{cyclostationary} \cite{Roychowdhury98}, i.e., the statistics will not be stationary, but will rather be periodic at the frequency of the drive signal.

The final element of the feedback loop is a phase shifter which is used to set the phase of the feedback so that it sustains the motion of the resonator counteracting the intrinsic dissipation as well as phase shifts deriving from the other parts of the feedback loop, we will suppose that there is a tunable component giving a total phase shift that can be tuned to select special operating points of the system.

\subsection{Closed loop equation of motion}

The effect of the feedback drive on the resonator may now be calculated using the envelope formalism by projecting the feedback signal and noise onto the dynamics near the carrier frequency -- the effect of other harmonics is made negligible by the strong filtering action of the high-Q resonator.
We introduce the complex amplitude of the drive $D(T)$ in analogy with Eq.~(\ref{signal}) through
\begin{equation}
\label{drive}
d(t)=\tfrac{1}{2}iD(T)e^{it}+\text{c.c.}+d_{1}(t),
\end{equation}
where the factor of $i$ is included so that real $D$ corresponds to positive feedback. The term $d_{1}(t)$ adding to the slow modulation of the basic oscillation, involves harmonics $e^{int},n\ne\pm 1$: these may not necessarily be small compared with the first term, since the amplifier may be strongly nonlinear, but will have a small effect on the resonator motion since they are are non resonant. $D(T)$ may be obtained from $d(t)$ using an integral analogous to Eq.~(\ref{A average})
\begin{equation}\label{D average}
D(T) =-\frac{i}{ \pi}\int_{\varepsilon^{-1}T- \pi}^{\varepsilon^{-1}T+ \pi}d(t)e^{-it}dt.
\end{equation}
Then
writing $b=\varepsilon\alpha$ and $c=\varepsilon\eta$, where the $\varepsilon$ factor reflects the weak nonlinearity,
and using the standard reduction procedure \cite{crossBook,LCreview} on Eq.~(\ref{resonatorEOM}), leads to the equation of motion for the complex amplitude of the form
\begin{equation}\label{ampEqClean}
    \frac{dA}{dT}+f(A)= \frac{1}{2}D,
\end{equation}
where
\begin{equation}\label{ampEqCleanResonator}
    f(A)=\frac{1}{2}A+\left(\frac{1}{8}\eta-\frac{3}{8}i\alpha|A|^{2}\right)A
\end{equation}
gives the intrinsic resonator terms, with $\alpha$ quantifying the strength of the nonlinear frequency pulling and $\eta$ the size of the nonlinear correction to the linear dissipation represented by the term $\tfrac{1}{2}A$. For feedback drive, the drive term is
\begin{equation}\label{ampEqCleanDrive}
    D(T)=\left[g(a)e^{i\Delta}+\Xi e^{i\Phi_N}\right]e^{i\Phi}.
\end{equation}
The first term in the braces gives the deterministic driving from the feedback with the real quantity $g(a)$, the strength of the driving, given by the magnitude of the fundamental harmonic of the output from the amplifier. The feedback will sustain the oscillations when the driving cancels the dissipation term proportional to $\dot q$, which occurs when the phase of the drive is near $\pi/2$ relative to the phase of $q$. The parameter $\Delta$ allows for a phase shift of the feedback relative to this value.
The second term in the braces gives the stochastic driving from noise $\Xi(T)=\Xi_{R}(T)+i\Xi_{I}(T)$ a complex stochastic noise acting on the slow time scale. It is convenient to define these noise components introducing a constant phase offset $\Phi_{N}$ from the resonator phase $\Phi$. The value of $\Phi_{N}$ will be chosen later to simplify the correlations of $\Xi_{R},\Xi_{I}$.
Note that $\Xi$ is defined relative to the phase $\Phi$ of the complex amplitude, which is dynamic on the slow time scale $T$. As we will explicitly demonstrate, the statistics of the noise $\Xi$ on the slow time scale are then \emph{stationary}, that is, $\langle\Xi_{\alpha}(T)\Xi_{\beta}(T')\rangle=C_{\alpha\beta}(T-T')$, reflecting the fact that there is no fixed time reference for a self sustained oscillator. The slow noise is therefore characterized by the spectra $S_{ij}(\Omega)$ (for $i,j$ either $R$ or $I$) defined by
\begin{equation}
\label{Xi_iXi_j}
\langle{\Xi}_{i}(\Omega){\Xi}_{j}(\Omega')\rangle=2\pi\varepsilon\delta(\Omega+\Omega')S_{ij}(\Omega),
\end{equation}
with $\Xi_{i}(\Omega)$ the Fourier transform
\begin{equation}
\label{Xi FT}
\Xi_{i}(\Omega)=\int_{-\infty}^{\infty}\Xi_{i}(T)e^{-i\Omega T}\,dT,
\end{equation}
and where the factor of $\varepsilon$ corresponding to the transformation to the slow time scale is included for convenience in the definition of $S_{ij}(\Omega)$.
To calculate the phase noise of the oscillator we need $S_{RR}(\Omega),S_{II}(\Omega)$ and the symmetric combination of the cross-correlation $S_{RI}^{s}(\Omega)=S_{RI}(\Omega)+S_{IR}(\Omega)$.

We can separate the noise into two components. The first component is noise in the feedback driving, typically arising from amplifier noise. The feedback noise results from various noise sources in the feedback circuit mixed with the periodic signal by the nonlinearity in the amplifier and any limiters in the loop, and so has complicated statistics. For a truly periodic signal in the loop the noise statistics is periodic rather than stationary. Nevertheless, we will show that the contribution to the slow noise $\Xi$ is stationary. The second noise component is from stochastic forces acting directly on the resonator: examples are thermomechanical noise associated with the dissipation of the resonator (analogous to Johnson noise in a resistor) and parameter noise such as fluctuations in the resonance frequency or dissipation coefficients.

\subsection{Amplifier gain function}

The amplifier gain function $g(a)$ together with the phase shift $\Delta$ are obtained by examining the drive $Q^{-1}d(t)$ on the resonator from the output of the amplifier-phase shifter system with a periodic input $a\cos t$, ignoring noise terms.
The drive is then projected onto the slow time scales using Eq.~(\ref{D average}) and the result is set equal to $g(a)e^{i\Delta}.$\footnote{In principle, the phase shift could depend on the input amplitude, but since this is an important control parameter of the oscillator system, we will assume that it is dominated by linear components, so that there is no important amplitude dependence.} We implement this calculation for a phenomenological model of the amplifier in \S\ref{phenomenological}. For a practical implementation the calculation would probably be done using circuit simulator models for the amplifier, or $g(a),\Delta,$ could be determined experimentally.

\subsection{Operating point}
\label{Sec: Operating point}

The first task is to find the operating point of the closed loop oscillator in the absence of noise as a function of the feedback phase, and the amplifier parameters.

We write the amplitude equation (\ref{ampEqClean}) for the oscillator, first without noise
\begin{equation}\label{ampEq}
   \frac{dA}{dT}+\left[1-\frac{1}{4}(3i\alpha-\eta)|A|^2\right]\frac{A}{2}=\frac{g(a)}{2}e^{i\Phi}e^{i\Delta}.
\end{equation}
Equation (\ref{ampEq}) separates into two real equations
\begin{eqnarray} \label{twoEquations}
  \frac{da}{dT} &=& -\frac{a}{2}\left(1+\frac{\eta}{4}a^2\right)+\frac{g(a)}{2}\cos\Delta=f_a(a), \\
  \frac{d\Phi}{dT} &=& \frac{3}{8}\alpha a^2+\frac{g(a)}{2a}\sin\Delta=f_{\Phi}(a),\nonumber
\end{eqnarray}
and the amplitude of oscillation, $a_0$, and the frequency $\Omega_0$, satisfy $f_a(a_0)=0$, $f_{\Phi}(a_0)=\Omega_0$.
From these equations we can find explicit results for $\Omega_{0}(a_{0})$ and $\Delta(a_{0})$
\begin{eqnarray}\label{freqAmp}
    \Omega_0&=&\frac{3\alpha a_0^2\pm\sqrt{16  g^2(a_{0})/a_0^2-\eta ^2a_0^4-8\eta a_0^2  -16 }}{8},\nonumber\\
    \Delta &=& \tan^{-1}\left(\frac{2\Omega_0-\frac{3\alpha a_0^2}{4}}{1+\frac{\eta a_0^2}{4}} \right).
\end{eqnarray}
These equations can be inverted numerically to give the operating point in terms of the feedback phase $a_{0}(\Delta),\Omega_{0}(\Delta)$ given the amplifier gain function $g(a)$.

\subsection{Phase noise}
\label{sec: oscillator phase noise}
The spectral output of a deterministic limit cycle is ideal, consisting of sharp peaks (delta functions) at the oscillator frequency and its harmonics. Such a system would be a perfect clock or frequency reference. The degradation of the performance is due to noise acting on the system. The wanderings of the phase variable from the ideal uniform progression $\omega_{0}t$ caused by the noise leads to a broadening of the spectral peaks. Often, this \emph{phase noise} is characterized by plotting the spectral density of the oscillator signal, on a log scale, as a function of the frequency offset $\omega-\omega_{0}$.

For frequency offsets small compared with the relaxation rate of perturbations returning to the limit cycle, typically of order $\omega_{0}/Q$, and for small noise amplitude, the phase noise can be calculated in terms of the projection of the noise along the \emph{phase sensitivity vector}  $\mathbf v_{\bot}$. The direction of the phase sensitivity vector may be related to the isochrons \cite{Winfree74,Guckenheimer75}, the surfaces in the phase space of the oscillator (here a curve in the two dimensional $a,\Phi$ space) such that all points on the surface asymptote to the same phase point on the limit cycle in the long time limit when the perturbation away from the limit cycle has decayed. The vector $\mathbf v_{\perp}$ is perpendicular to the direction of the isochron at the limit cycle. The phase sensitivity vector is also the zero-eigenvalue adjoint eigenvector of the linearized time evolution near the fixed point giving the oscillator state \cite{DemirMehrotra00}, and in this approach is often called the \emph{perturbation projection vector}.

In the complex amplitude formulation, the result may be derived as follows. Equation (\ref{ampEqClean}), with the complex amplitude represented in magnitude-phase form by the vector $\mathbf x=(a,\Phi)$, may be written as
\begin{equation}\label{X EOM}
\frac{d\mathbf x}{dT}=\mathbf f(a)+\Xi_{R}\mathbf v_{R}+\Xi_{I}\mathbf v_{I},
\end{equation}
where $\mathbf f(a)=(f_{a}(a),f_{\Phi}(a))$ and the noise vectors
\begin{eqnarray}\label{noiseVectors}
  \textbf{v}_{R} &=& \frac{1}{2}\left(\cos\Phi_{N},\frac{\sin\Phi_{N}}{a_0}\right),\\
  \textbf{v}_{I} &=& \frac{1}{2}\left(-\sin\Phi_{N},\frac{\cos\Phi_{N}}{a_0}\right),\nonumber
\end{eqnarray}
define the coupling of the two components of noise to the system. Equation (\ref{X EOM}) corresponds to adding the stochastic terms to Eqs.~(\ref{twoEquations}).
The phase noise is calculated by linearizing Eqs.~(\ref{X EOM}) in the small noise.
For small frequency offsets, the $da/dT$ term in the magnitude component of the linearized Eqs.~(\ref{X EOM}) can be neglected, giving an explicit equation for the magnitude fluctuations in terms of the noise. Inserting these into the phase fluctuation equation then gives a single stochastic equation for the phase evolution \cite{kenig12}
\begin{equation}
    \dot{\phi}=P_R\,\Xi_{R}+ P_I\,\Xi_{I},\label{stochastic phase}
\end{equation}
where $\phi=\Phi-\Omega_0T$, and the constants $P_{R},P_{I}$
\begin{equation}
\label{phaseSensitivity}
P_{R}=\mathbf v_{R}\cdot\mathbf v_{\bot},\quad P_{I}=\mathbf v_{I}\cdot\mathbf v_{\bot},
\end{equation}
are the noise projections along the phase sensitivity vector given by
\begin{equation}\label{zeroMode}
    \textbf{v}_\bot=\left(-\frac{f_{\Phi}'(a_{0})}{f'_a(a_{0})},1\right).
\end{equation}
 More formally, the methods of Ref.~\cite{Demir02} may be used to derive these results, see Appendix \ref{Appendix: phase equation}.
A key simplification of our approach is that $\mathbf{v}_{\perp}$ and the noise vectors $\mathbf{v}_{R},\mathbf{v}_{I}$ are \emph{constant} vectors \cite{kenig12}. This comes from the fact that the oscillator phase $\Phi$ does not appear on the right hand side of Eq.~(\ref{X EOM}) in either the deterministic or stochastic terms, ultimately deriving from the phase symmetry of the description in terms of the complex amplitude.

We have written the vectors in terms of two components $(v_{a},v_{\Phi})$ giving the magnitude and phase coordinates. A more intuitive representation is given by defining vectors in the two dimensional phase space of the limit cycle. Vectors $\mathbf{v}$, such as $\mathbf{v}_{R,I}$, are given,  in polar form
in this space, by multiplying the phase component by $a_{0}$: $\mathbf{v}\to v_{a}\hat{\mathbf{a}}_{0}+a_{0}v_{\Phi}\hat{\boldsymbol\Phi}_{0}$, where $\hat{\mathbf{a}}_0,\hat{\boldsymbol\Phi}_0$ are unit vectors in the magnitude and phase directions at the point on the limit cycle. \emph{Adjoint} vectors $\mathbf{v}^{\dagger}$, such as the phase sensitivity vector $\mathbf{v}_{\perp}$, are given by \emph{dividing} the phase component by $a_{0}$: $\mathbf{v}^{\dagger}\to v_{a}^{\dagger}\hat{\mathbf{a}}_{0}+a_{0}^{-1}v_{\Phi}^{\dagger}\hat{\boldsymbol\Phi}_{0}$. This preserves the scalar products $\mathbf{v}^{\dagger}\cdot\mathbf{v}$. In this representation, the vectors $\mathbf{v}_{R,I},\mathbf{v}_{\perp}$ all rotate at the rate $\Omega_{0}$ together with the point on the limit cycle. Of course, the scalar products $P_{R},P_{I}$ giving the noise projections remain time independent.

The oscillator phase given by Eq.~(\ref{stochastic phase}) is a stochastic process quantified by the variance $V(\tau)=\langle[\delta\phi(T+\tau)-\delta\phi(T)]^2\rangle$ with $\delta\phi(T)=\phi(T)-\langle\phi(T)\rangle$.\footnote{There are corrections to the drift frequency  also proportional to the noise strength that give a small shift of the oscillator spectral peaks that we do not address here.}
This variance can be calculated by Fourier transforming Eq.~(\ref{stochastic phase})
\begin{equation}\label{phaseVariance}
    V(\tau)=\frac{4\varepsilon}{\pi}\sum_{i,j=R,I}P_iP_j\int_{0}^{\infty}S_{ij}(\Omega)\left[\frac{\sin(\Omega\tau/2)}{\Omega}\right]^2d\Omega.
\end{equation}
A simple common case (see below) is if the spectrum of the slow noise is white, $S_{ij}(\Omega)=F_{ij}$ independent of $\Omega$: in this case the variance grows linearly in time
\begin{equation}
\label{variance white noise}
V(\tau)=\varepsilon\sum_{i,j=R,I}P_iP_jF_{ij}\,\tau,
\end{equation}
corresponding to a random walk of the phase or \emph{phase diffusion}.

The spectrum of the oscillator is the Fourier transform of the autocorrelation function of the output of the oscillator which we take to be the displacement of the resonator $q(t) = a \cos(t + \Phi)$. Neglecting amplitude fluctuations and after transients have died out, the spectral density of the displacement can be written as $S(\omega)=a^2_{0}[\bar S(\omega+\bar\omega_{0})+\bar S(\omega-\bar\omega_{0})]/4$ with
\begin{equation}
\label{phase noise fourier}
\bar S(\omega)=\mathcal F[e^{-\tfrac{1}{2}V(t/Q)}],
\end{equation}
and $\bar\omega_{0}$ is the scaled oscillation frequency.
The well known Leeson expression \cite{Leeson66} for the phase noise spectrum results from evaluating $\bar S(\omega)$ away from the carrier frequency where the Fourier transform in Eq.~(\ref{phase noise fourier}) is dominated by small times for which the variance $V(\tau)$ is small.\footnote{More complete expressions for the noise spectra in other limits are discussed in Ref.~\cite{kenig12}.} In this case the exponential can be expanded to first order, so that for $\Omega$ not too small
\begin{equation}\label{phase noise spectrum}
\bar S(\varepsilon\Omega)=
\frac{\sum_{i,j=R,I}P_iP_jS_{ij}(\Omega)}{\Omega^{2}}.
\end{equation}
Equation (\ref{phase noise spectrum}) together with expressions for the slow noise spectra $S_{ij}(\Omega)$ provide a complete prescription for calculating the phase noise of oscillators in the regimes usually of interest.
It reproduces the standard result \cite{Leeson66} for the oscillator phase noise as a function of the offset frequency $\omega_{m}$, namely an $\omega_{m}^{-2}$ dependence for white noise sources ($S_{ij}(\Omega)=\text{constant}$), $\omega_{m}^{-3}$ for 1/f noise sources ($S_{ij}(\Omega)\propto\Omega^{-1}$), etc., and provides a simple route to a quantitative calculation.
The phase noise is conventionally quoted as

\begin{equation}\label{pn}
    L(\omega_m)=10\log_{10}\left[\bar S\left(\frac{\omega_{m}}{\omega_0}\right)\frac{1}{\omega_0}\right]
\end{equation}
in dBc/Hz \cite{kenig12,Demir02}.

\section{Calculating the slow noise}

The slow noise $\Xi(T)$ is given in terms of the noise $\xi(t)$ in the drive function $d(t)$ by averaging over one period as in Eq.~(\ref{D average}). The real and imaginary components are given by
\begin{eqnarray}\label{cartezian noise}
  \Xi_R(T) &=&\frac{1}{ \pi}\int_{\varepsilon^{-1}T- \pi}^{\varepsilon^{-1}T+ \pi}\xi(t)\cos(t+\psi_N)dt,\\
  \Xi_I(T) &=& -\frac{1}{ \pi}\int_{\varepsilon^{-1}T- \pi}^{\varepsilon^{-1}T+ \pi}\xi(t)\sin(t+\psi_N)dt,\nonumber
\end{eqnarray}
with $\psi_N=\Phi+\Phi_{N}+\pi/2$ the phase deriving from the phase factor used in defining $\Xi$ in Eq.~(\ref{ampEqCleanDrive}) and the factor of $i$ introduced in Eq.~(\ref{drive}).
From these expressions the Fourier transforms $\Xi_{R}(\Omega),\Xi_{I}(\Omega)$ and then the correlations $\langle\Xi_{i}(\Omega)\Xi_{j}(\Omega')\rangle$ and hence the spectral densities of the slow noise $S_{ij}(\Omega)$ can be obtained.
We now evaluate these spectral densities for feedback and resonator noise sources.

\subsection{Feedback noise}
\label{Feedback noise}

As described in \S\ref{subsec: Basic setup}, to calculate the noise generated by the amplifier (and other components that might be in the circuit such as a phase shifter and a limiter) it is sufficient to consider a \emph{periodic} input signal $a\cos(t+\Phi)$, neglecting the slow time dependence of $a$ and $\Phi$. The noise $\xi(t)$ is then cyclostationary, with statistics that are periodic.
Following the approach of Roychowdhury and Long \cite{Roychowdhury98} the correlation function of the noise is expressed in the form
\begin{equation}
\label{noise correlation}
\langle\xi(t)\xi(t')\rangle=\sum_nR_{n}(t-t')e^{int},
\end{equation}
and then
\begin{equation}
\label{R-Q}
R_{n}(\tau)=\int\frac{d\omega}{2\pi}Q_{n}(\omega)e^{i\omega\tau},
\end{equation}
with $Q_{n}(\omega)$ the harmonic power spectral densities (HPSDs). For stationary noise we would have $Q_{n\ne 0}=0$, and so nonzero values of these quantities demonstrate the cyclostationary nature of the noise.
The $Q_{n}$ satisfy the symmetry relations
\begin{equation}\label{Qsymmetry}
    Q_{n}(-\omega)=Q_{n}(\omega-n),\quad Q_{-n}(-\omega)=Q_{n}^{*}(\omega).
\end{equation}

We consider a single stationary noise source $\xi_s(t)$ with spectrum given by the Fourier transform of the correlation function $Q_{s_0}(\omega)={\cal F}[\langle\xi_s(t)\xi_s(0)\rangle]$. This noise goes through the various  feedback components and transforms to the noise $\xi(t)$ in the feedback drive $d(t)$. Since both noises $\xi(t),\xi_{s}(t)$ are assumed to be small perturbations, they are related through the linear response function of the time-varying system between the noise source and the output of the feedback system
\begin{equation}\label{output noise}
    \xi(t)=\int_{-\infty}^{\infty}h(t,t')\xi_s(t')dt',
\end{equation}
where $h$ is a periodic function with the same periodicity as the input signal to the amplifier \cite{Roychowdhury98}, i.e.\ $h(t+2n\pi,t'+2n\pi)=h(t,t')$ for any integer $n$, and will in general depend in a nonlinear way on the input signal. The quantity $h(t,t')$ is now written in terms of harmonic transfer functions (HTFs) of the feedback system. It is useful to define these for zero input phase $\Phi=0$, and then the response for a general phase $\Phi$ is given by a time translation. Thus we write
\begin{equation}\label{corrr}
     h(t,t')=\sum_n\,e^{in\Phi}h_n(t-t')e^{int},\\
\end{equation}
with the zero-phase harmonic transfer functions $H_{n}(\omega)={\cal F}[h_{n}(t)]$ satisfying $H_{-n}(-\omega)=H_{n}^{*}(\omega)$ since $h(t,t')$ is real. Following Ref.~\cite{Roychowdhury98}, the HPSDs of the output noise Eq.~(\ref{output noise}) are related to the stationary spectrum of the noise source through
\begin{equation}\label{spectrumHarmonics}
    Q_l(\omega)=e^{il\Phi}\sum_{n}H_n(-\omega-n)Q_{s_0}(\omega+n)H_{l-n}(\omega+n).
\end{equation}

The slow noise spectra $\langle{\Xi}_{i}(\Omega){\Xi}_{j}(\Omega')\rangle$ are evaluated from Eqs.~(\ref{cartezian noise}) in terms of $Q_{l}$. This is done in Appendix \ref{Appendix: slow noise} using the fact that $\varepsilon$ is small. There we derive Eq.~(\ref{Xi_iXi_j}),
showing that the slow noise is indeed stationary as already mentioned, and calculate results for the spectral densities $S_{ij}(\Omega)$ in terms of the spectrum $Q_{s_{0}}(\omega)$ of the noise source and the HTFs $H_{n}$ of the amplifying system. To calculate the phase noise of the oscillator Eq.~(\ref{phaseVariance}) or Eq.~(\ref{phase noise spectrum}) we need $S_{RR}(\Omega),S_{II}(\Omega)$ and the symmetric combination of the cross-correlations $S_{RI}^{s}(\Omega)=S_{RI}(\Omega)+S_{IR}(\Omega)$.
As described in the Appendix, the results for these quantities fall into two classes depending on the nature of the noise sources.

\subsubsection{Broadband noise}

The spectrum of many noise sources, such as Johnson noise of resistors, will not have structure on the $O(\varepsilon)$ frequency scales corresponding to the width of the response of the resonator. We call these sources broadband. For these sources we derive the results Eqs.~(\ref{specs broadband appendix}) in the Appendix
\begin{eqnarray}
\label{specs broadband}
S_{RR}(\Omega)&=&2\big\{\sum_{n}Q_{s_{0}}(n)|H_{n-1}(-n)|^{2}\\
&+&\textmd{Re}[e^{2i\bar\psi_N}\sum_{n}Q_{s_{0}}(n)H_{n-1}(-n)H_{n+1}^{*}(-n)]\big\},\nonumber\\
S_{II}(\Omega)&=&2\big\{\sum_{n}Q_{s_{0}}(n)|H_{n-1}(-n)|^{2}\nonumber\\
&-&\textmd{Re}[e^{2i\bar\psi_N}\sum_{n}Q_{s_{0}}(n)H_{n-1}(-n)H_{n+1}^{*}(-n)]\big\},\nonumber\\
S_{RI}^{s}(\Omega)&=&-4\textmd{Im}[e^{2i\bar\psi_N}\sum_{n}Q_{s_{0}}(n)H_{n-1}(-n)H_{n+1}^{*}(-n)],\nonumber
\end{eqnarray}
with $\bar\psi_N=\Phi_{N}+\tfrac{\pi}{2}$ and $\textmd{Re},\textmd{Im}$ denoting real and imaginary parts. Note that there is no dependence on $\Omega$ on the right hand sides of Eqs.~(\ref{specs broadband}) so that the noise is white on the frequency scale of the slow time dependence.

In defining the noise $\Xi$ in Eq.~(\ref {ampEqCleanDrive}) we included a reference phase $\Phi_{N}$ to be chosen to simplify the correlations of the slow noise. In particular we choose $\Phi_N$ to eliminate the cross correlation $S_{RI}^{s}$.
The resulting expression for the slow noise is white with uncorrelated quadratures with spectra
\begin{eqnarray}
\label{specs broadband uncorrelated}
&S_{RR}=2\big\{\sum_{n}Q_{s_{0}}(n)|H_{n-1}(-n)|^{2}\\
&\pm|\sum_{n}Q_{s_{0}}(n)H_{n-1}(-n)H_{n+1}^{*}(-n)|\big\},\nonumber\\
&S_{II}=2\big\{\sum_{n}Q_{s_{0}}(n)|H_{n-1}(-n)|^{2}\nonumber\\
&\mp|\sum_{n}Q_{s_{0}}(n)H_{n-1}(-n)H_{n+1}^{*}(-n)|\big\},\nonumber
\end{eqnarray}
where the two possible sign choices, resulting from choices of $\Phi_N$ differing by $\pi/2$, correspond to the arbitrary choice of which of two orthogonal directions to assign to `$R$' and which to `$I$'.

The expressions (\ref{specs broadband uncorrelated}) have the intuitive interpretation that the slow noise (i.e.\ the noise near the carrier frequency) is given by the noise source intensity at various harmonics of the carrier frequency mixed up or down to the vicinity of the carrier with a strength depending on the HTFs of the amplifier. On the other hand, a completely linear amplifier would generate stationary feedback noise, giving the slow noise intensities
\begin{equation}\label{stationary}
S_{RR}=S_{II}=2Q_{s_{0}}(1)|H_0(1)|^2.
\end{equation}
We calculate the phase noise for this case in \S\ref{sec:linear amp}.

The slow noise strengths can be reduced if it is possible to put a filter, filtering the signal around the carrier frequency, between noise generating but linear early stages of the amplifier and later nonlinear stages. In this case only the $n=\pm1$ components get through to the nonlinear stages where up and down conversion of the noise occurs. In this case (we call ``filtered white noise'') we find
\begin{eqnarray}
\label{specs filtered uncorrelated}
&S_{RR}=2Q_{s_{0}}(1)(|H_0(1)|\pm|H_{2}(-1)|)^2,\\
&S_{II}=2Q_{s_{0}}(1)(|H_0(1)|\mp|H_{2}(-1)|)^2.\nonumber
\end{eqnarray}
This type of noise was analyzed in Ref.~\cite{wiesenfeld13}.

\subsubsection{1/f noise}
\label{sec: 1/f general}

The second class of noise sources important to consider are those with intensity $Q_{s0}(\omega)$ growing as the frequency $f=\omega/2\pi$ decreases. Many amplifiers show such noise with an intensity often growing at low frequencies as a power law $f^{-\nu}$ with $\nu$ close to unity. Such noise is typically described as 1/f noise. For this type of noise source we derive the results Eqs.~(\ref{specs 1/f appendix}) in the Appendix
\begin{eqnarray}\label{specs 1/f}
S_{RR}(\Omega)&=&2Q_{s_{0}}(\varepsilon\Omega)|H_{1}(0)|^{2}\{1+\cos[2(\phi_{H}-\bar\psi_N)]\},\nonumber\\
S_{II}(\Omega)&=&2Q_{s_{0}}(\varepsilon\Omega)|H_{1}(0)|^{2}\{1-\cos[2(\phi_{H}-\bar\psi_N)]\},\nonumber\\
S_{RI}^{s}(\Omega)&=&4Q_{s_{0}}(\varepsilon\Omega)|H_{1}(0)|^{2}\sin[2(\phi_{H}-\bar\psi_N)].
\end{eqnarray}
where $\phi_{H}$ is the phase of $H_{1}(0)$, i.e., $H_{1}(0)=H_{-1}^{*}(0)=|H_{1}(0)|e^{i\phi_{H}}$. We now make the cross correlation $S_{RI}^{s}$ zero by choosing $\bar\psi_{N}=\phi_{H}$ giving $\Phi_{N}=\phi_{H}-\tfrac{\pi}{2}$. With this choice of reference direction the two quadratures of the noise have the spectra
\begin{equation}\label{specs 1/f uncorrelated}
    S_{RR}(\Omega)=4Q_{s_{0}}(\varepsilon\Omega)|H_{1}(0)|^{2},\quad S_{II}(\Omega)=0.
\end{equation}
Equation (\ref{specs 1/f uncorrelated}) shows that for a single 1/f noise source, the slow noise lies along a \emph{line} in the complex amplitude space (making an angle $\Phi_N$ to the vector of the feedback signal) rather than filling out a ball as for broadband noise sources. This has the important consequence of the potential to eliminate the resulting phase noise by tuning the phase sensitivity vector to be perpendicular to this direction \cite{kenig1/f}. We demonstrate this for the phenomenological amplifier model in \S\ref{sec:1/f} below. Also note that the slow noise inherits the spectrum $Q_{s_{0}}(\varepsilon\Omega)$ of the noise source, which is up-converted from near zero to near the carrier frequency (through the harmonic transfer function $H_1$), as shown by measurements of amplifier phase noise in \cite{amplifierNoiseReview}.

The 1/f noise leading to Eqs.~(\ref{specs 1/f uncorrelated}) will dominate close to the carrier frequency (small $\Omega$), but as $\Omega$ increases this contribution may become smaller than terms analogous to the $n\ne 0$ terms of Eq.~(\ref{specs broadband uncorrelated}) -- see Eqs.~(\ref{specs 1/f correction appendix}) in the Appendix. Far enough away from the carrier where these terms dominate we may choose a new value of $\Phi_{N}$ to make $S_{RI}^{s}$ in Eqs.~(\ref{specs 1/f correction appendix}) zero, and then the slow noise spectra become
\begin{eqnarray}
\label{1/f noise tail}
&S_{RR}(\Omega)=2\big\{\sum_{n\ne 0}Q_{s_{0}}(n)|H_{n-1}(-n)|^{2}\\
&\pm|\sum_{n\ne 0}Q_{s_{0}}(n)H_{n-1}(-n)H_{n+1}^{*}(-n)|\big\},\nonumber\\
&S_{II}(\Omega)=2\big\{\sum_{n\ne 0}Q_{s_{0}}(n)|H_{n-1}(-n)|^{2}\nonumber\\
&\mp|\sum_{n\ne 0}Q_{s_{0}}(n)H_{n-1}(-n)H_{n+1}^{*}(-n)|\big\},\nonumber
\end{eqnarray}
as in Eq.~(\ref{specs broadband uncorrelated}) without the $n=0$ terms.
Thus the slow noise crosses over from 1/f near the carrier to white further away from the carrier.

\subsection{Resonator noise}

Two types of noise acting directly on the resonator are expected. Firstly, there will be an additive noise force term in Eq.~(\ref{resonatorEOM}) related to the linear dissipation term via the fluctuation-dissipation theorem. This is thermomechanical noise for a mechanical resonator and Johnson noise for an electronic resonator. The spectrum is usually white\footnote{The spectrum is non-white in the quantum regime $\omega\gtrsim k_{B}T/\hbar$.} and the noise intensity is proportional to the temperature and the dissipation coefficient. There may also be additive noise associated with the nonlinear dissipation. The second type of noise is parameter fluctuations. For example the mass of a mechanical resonator may fluctuate due to gas molecules binding and unbinding from the structure, and the stiffness may fluctuate due to temperature fluctuations. The spectra of these noises may be white, white filtered by a response of the device (e.g. thermal fluctuations will be quenched above a time scale determined by the thermal contact to the environment), or 1/f. Since the $\ddot q$ ``mass'' term and the $q$ ``spring constant'' term are the largest terms in the equation of motion (\ref{resonatorEOM}), fluctuation of these coefficients are likely to be most important, and we will focus on these, although other fluctuations are easily included by analogous methods.

\subsubsection{Additive noise}

Averaging an additive noise $\varepsilon\xi(t)$
over a period as in Eq.~(\ref{cartezian noise}) to obtain the slow noise $\Xi(T)$ corresponds to sampling the noise near the carrier frequency. The resulting noise is white on the slow time scale and isotropic
\begin{align}
S_{RR}(\Omega)&=S_{II}(\Omega)=2Q_{s0}(1),\\
S_{RI}^{s}(\Omega)&=0, \nonumber
\end{align}
with $Q_{s0}(\omega)$ the spectrum of $\xi(t)$. For thermodynamic noise, such as thermomechanical noise, the noise strength is related to the dissipation coefficient via the fluctuation-dissipation theorem. For a mechanical resonator, this gives
\begin{equation}
Q_{s0}(\omega)=2Q\frac{k_{B}T}{K},
\end{equation}
with $K=m\omega_{0}^{2}$ the stiffness constant. A similar result applies to an electronic LCR resonator with the capacitance replacing the stiffness constant.

\subsubsection{Parameter noise}

For a fluctuating mass $m\to m(1+\varepsilon\xi_{m}(t))$ there will be an additional stochastic drive term $-\varepsilon\xi_{m}(t)\ddot q$ on the right hand side of Eq.~(\ref{cartezian noise}). Similarly, for a fluctuation spring constant $K\to K(1+\varepsilon\xi_{K}(t))$ there will be a stochastic drive term $-\varepsilon\xi_{K}(t)q$. As in calculating the feedback noise, to leading order in the noise strength and the small parameter $\varepsilon$ we may evaluate these source terms neglecting the slow time dependence of the amplitude $A(T)$, leading to noise source terms $\varepsilon a\cos(t+\Phi)\xi_{m}$ and $-\varepsilon a\cos(t+\Phi)\xi_{K}$. These are of the same form, and so we need to calculate the slow noise from a noise source $\varepsilon a\cos(t+\Phi)\xi_s(t).$\footnote{For independent mass and spring constant fluctuations we add the resulting phase diffusion; if the two noises are correlated, for example both resulting from a temperature fluctuation, we add the noise amplitudes before calculating the phase diffusion.}

We could proceed by evaluating the integrals in Eq.~(\ref {cartezian noise}) etc., but it is easier to recognize the slow noise from the multiplicative parameter noise as being equivalent to the slow noise from an ideal square-law mixer, and then using the formalism of \S\ref{Feedback noise}. The only nonzero components of the harmonic transfer functions of such a mixer with input signal $a\cos t$ are $H_{1}(\omega)=H_{-1}(\omega)=\tfrac{1}{2}a$. Then the choice $\Phi_{N}=0$ leads to the expressions for the slow noise spectra
\begin{align}\label{parameterNoise}
S_{RR}(\Omega)&= \tfrac{1}{2}a^{2}Q_{s0}(2),\\
S_{II}(\Omega)&= a^{2}[Q_{s0}(\varepsilon\Omega)+\tfrac{1}{2}Q_{s0}(2)],\nonumber\\
S_{RI}^{s}(\Omega)&= 0,\nonumber
\end{align}
where $Q_{s0}$ is the spectral density of $\xi_s(t)$. The $\Omega$ dependence on the right hand side can be ignored except for 1/f noise when the first term in $S_{II}$ dominates.
Notice that the up-conversion of the low frequency noise source $Q_{s0}(\Omega\simeq 0)$ leads to noise purely in the phase direction $S_{II}$, as might be expected for noise leading to fluctuations in the resonance frequency of the resonator. This is the case for a 1/f noise source. However, for a broadband noise source, the down conversion of the noise near twice the carrier frequency $Q_{s0}(2)$ leads to an additional isotropic contribution. For white noise sources $Q_{s0}(\omega)=f_{0}$ the noise is predominantly along the phase direction $S_{II}=3S_{RR}=\tfrac{3}{2}a^{2}f_{0}$.

Fluctuations in the other parameters of the resonator can be treated analogously. Note that fluctuations in the dissipation coefficient $\gamma\to\gamma(1+\xi_{\gamma})$ will lead to a stochastic force $\varepsilon a\sin(t+\Phi)\xi_{\gamma}$, with a $\pi/2$ phase shift from the mass or stiffness constant fluctuations. The results for $S_{RR},S_{II}$ will correspondingly be interchanged, so that the noise is predominantly along the magnitude quadrature, as expected physically.

\section{Strategies for reducing the phase noise}

We now discuss strategies to reduce the oscillator phase noise using the feedback phase to tune the oscillator to operating points where the sensitivity to particular noise sources is reduced or eliminated. An important ingredient that allows this approach is that the oscillator frequency $\Omega_{0}$, given by setting $d\Phi/dT=\Omega_{0}$ in Eq.~(\ref{twoEquations}), depends on the feedback phase $\Delta$ through both terms in $f_{\phi}$: the first term describes the nonlinear frequency dependence of the resonator, and depends on $\Delta$ through the dependence of the oscillation amplitude on this parameter; the second term gives the direct dependence of the oscillator frequency on the feedback phase, present even for a linear resonator.

As we have demonstrated in the previous section, a single  noise source will in general lead to a simple frequency dependence of the slow noise spectra appearing in Eqs.~(\ref{phaseVariance},\ref{phase noise spectrum}) for the oscillator phase noise: broadband noise sources lead to constant spectra, and a noise source with intensity increasing at low frequencies as $f^{-\nu}$ will, when upconverted by nonlinear processes, lead to slow noise spectra varying in the analogous way $S_{ij}(\Omega)\propto\Omega^{-\nu}$.

In general the oscillator phase noise will result from the combination of many different noise sources, perhaps with different spectra, leading to complicated frequency dependences $S_{ij}(\Omega)$ for the total noise. It is not likely that tuning the single parameter $\Delta$ will lead to a strong suppression of the noise Eq.~(\ref{phase noise spectrum}) over a significant frequency range in this situation.

If however the oscillator noise is dominated by a single noise source,
or at least by noise sources with the same spectra over some range of interest, the frequency dependence on the right hand side of Eqs.~(\ref{phaseVariance},\ref{phase noise spectrum}) will be common to all the terms in the sum and can be factored out. Furthermore, the value of $\Phi_{N}$ can be chosen to make the cross-correlation $S_{RI}^{s}$ zero, so that the noise forms a (noncircular) ball in the two-dimensional complex amplitude space, with independent fluctuations of strengths $S_{RR}(\Omega),S_{II}(\Omega)$ along and perpendicular to the direction specified by $\Phi_{N}$. The oscillator phase noise is then the sum of the effects of these two independent noises, given by their projections along the noise sensitivity vector $\mathbf{v}_{\perp}$.

It is useful to separate the geometric characteristics of the noise, given by the shape and orientation of the noise ball, from the overall spectrum. To do this we define the total noise spectrum $S_{\text{tot}}=S_{RR}+S_{II}$ and  the effective phase sensitivity $P^2_{\textmd{eff}}(\Delta)$ given by the weighted combination of $P_{R}^{2}$ and $P_{I}^{2}$
\begin{equation}
\label{P_eff}
P^2_{\text{eff}}(\Delta)=\frac{S_{RR}}{S_{\text{tot}}}P_{R}^{2}+\frac{S_{II}}{S_{\text{tot}}}P_{I}^{2}.
\end{equation}
Note that $P^2_{\text{eff}}$ is independent of frequency over frequency ranges for which $S_{RR},S_{II}$ have the same frequency dependence.
Using this expression, Eq.~(\ref{phase noise spectrum}) for example becomes
\begin{equation}
\label{Separated noise spectrum}
\bar S(\varepsilon\Omega)=\frac{P_{\text{eff}}^{2}(\Delta)S_{\text{tot}}(\Delta,\Omega)}{\Omega^{2}}.
\end{equation}
It should be noted that $S_{\text{tot}}$ may depend implicitly on the feedback phase $\Delta$. For example, the up and down conversion of noise by mixing with the signal in the amplifier depends on the magnitude of the input signal to the amplifier determined by the resonator amplitude $a_{0}(\Delta)$. Thus the full noise optimization with respect to the feedback phase must be performed on the whole expression Eq.~(\ref{Separated noise spectrum}). However, the focus of our work is to use the nonlinear behavior of the resonator to reduce the phase noise, exploiting the phase space geometry of the noise forces and the phase sensitivity of the resonator expressed by $P_{\text{eff}}$.
This behavior can be conveniently displayed by plotting the quantity $10\log_{10}\left[P^2_{\textmd{eff}}(\Delta)\right]$, corresponding to the conventional way of describing the phase noise, Eq.~(\ref{pn}).

Usually both $S_{RR},S_{II}$ will be nonzero so that the noise perturbations form a ball in phase space. In this case, it will typically not be possible to eliminate the effects of both components of the noise by tuning the oscillator to a special operating point. However, if one component of the noise is zero, so that the noise perturbations lie along a line $\mathbf v_{n}$ in phase space, it will often be possible to eliminate the effect of the remaining noise on the oscillator phase by tuning the parameters so that the direction of this line is orthogonal to the phase sensitivity direction, $\mathbf v_{\bot}\cdot\mathbf v_{n}=0\Rightarrow P_{\text{eff}}=0$. We first explore this possibility of complete noise elimination, and then strategies for noise reduction when this is not possible.

\subsection{Complete noise elimination}

We can identify two situations where the complete elimination of a noise source is possible.

\subsubsection{Feedback noise with a saturated amplifier}

For a saturated amplifier, where the magnitude of the output is independent of the input, or if a limiter is included in the circuit after the amplification stage, the noise in the magnitude quadrature of the feedback is suppressed, and the noise is purely in the feedback phase quadrature: with the choice $\Phi_{N}=\Delta$ we get $\Xi_{R}=0$. Thus only the projection $P_{I}=\mathbf{v}_{I}\cdot\mathbf{v}_{\perp}$ is relevant to the phase noise.
Using Eqs.~(\ref{twoEquations},\ref{noiseVectors},\ref{phaseSensitivity}), explicit calculation shows that the phase sensitivity to amplifier noise in the phase quadrature can be directly related to the dependence $\Omega_{0}(\Delta)$ of the oscillator frequency on the feedback phase \cite{wiesenfeld13}
\begin{eqnarray}\label{PI}
    P_I=\frac{1}{g(a_{0})}\frac{d\Omega_0}{d\Delta}.
 \end{eqnarray}
This result has the intuitive interpretation that noise in the phase quadrature is equivalent to taking fluctuations in the phase-shift parameter $\Delta$. Equation (\ref{PI}) shows that this noise has no effect on the phase diffusion for values of the phase shift $\Delta$ for which the frequency is insensitive to $\Delta$, and the condition for eliminating the effect of feedback phase noise is $d\Omega_0/d\Delta=0$.
This result generalizes the proposal of Greywall et al.\ \cite{greywall,YurkePra} where they showed through both theory and measurements that the effect of the amplifier noise $\Xi_{I}$ on the oscillator phase noise could be eliminated by choosing a feedback level and phase so that the resonator is driven exactly at the Duffing critical point where the amplitude-frequency and phase-frequency curves of the driven resonator become nonmonotonic. This generalized principle was also applied previously in \cite{driscoll} to eliminate amplifier noise in a quartz crystal oscillator. The condition was reinterpreted and generalized in a number of papers \cite{kenig12,wiesenfeld13,Villanueva13,kenig1/f}.

\subsubsection{1/f noise}

In \S\ref{sec: 1/f general} we showed that a single dominant 1/f noise source in the feedback loop or in a parameter of the resonator leads to slow noise along a line $\mathbf v_{n}$ in phase space, instead of forming a ball. This leads to the potential for elimination by finding parameters for which $\mathbf v_{\bot}\cdot\mathbf v_{n}=0$ \cite{kenig1/f}. For parameter noise in the mass or stiffness constant leading to fluctuations in the resonance frequency, $\mathbf{v}_{n}$ is along the phase direction of the resonator motion.
Noise in this quadrature has a direct effect on the oscillator phase noise (note that the component of $\mathbf{v}_{\perp}$, Eq.~(\ref{zeroMode}), in this direction is unity), and cannot be reduced by adjusting the feedback phase.  Noise in the dissipation coefficient leads to slow noise in the magnitude direction, and the resulting phase noise is eliminated by tuning to the operating point where $\partial f_\Phi/\partial a=0$ (see Eq.~(\ref{twoEquations})). Correlated fluctuations in the parameters, for example due to temperature fluctuations of the device may lead to a $\mathbf{v}_{n}$ in some general direction, and it may be possible to eliminate the phase noise by tuning to where $\mathbf{v}_{\perp}$ is perpendicular to this direction. For feedback noise, both $\mathbf v_{\bot}$ and $\mathbf v_{n}$ depend on the phase shift $\Delta$, and the possibility of tuning $\mathbf v_{\bot}\cdot\mathbf v_{n}$ to zero for this noise depends on the detailed characteristics of the amplifier. We investigate this for a phenomenological amplifier model in \S\ref{phenomenological}.

\subsection{Incomplete noise quenching}
\label{Incomplete}

Some useful noise quenching can still be obtained in the general case where $\Xi_{R},\Xi_{I}$ are both nonzero and imperfectly correlated, so that the noise perturbations form a ball in phase space.

For feedback noise, the best way to do this will depend on the details of the amplifier configuration, and so a calculation of $P^2_{\textmd{eff}}(\Delta)$ is required. One generic approach is to move the amplifier operation point to a more saturated condition, which will tend to reduce the noise component along the magnitude of the feedback, and then to use the nonlinear properties of the resonator to reduce the effects of the other noise quadrature. Useful insights on this approach can be obtained from a phenomenological model of an amplifier described by a nonlinear gain function  discussed in the following section.

For the direct resonator noise and the noise sources most likely to be important, we found that $\Phi_{N}=0$, so that the noise along the magnitude and phase quadratures of the resonator motion are uncorrelated. These noise sources will be independent of the feedback phase. Again, the noise in the phase quadrature directly adds to the oscillator phase noise and cannot be eliminated by adjusting the feedback phase, although the effect is reduced by going to larger oscillation amplitudes. Thus the best that can be done is to tune the feedback phase to eliminate the effect of the magnitude quadrature of the noise, i.e.\ to eliminate amplitude-phase noise conversion.  This occurs where $\partial f_\Phi/\partial a=0$. Note that this condition involves the cancellation of the contributions from the two terms in $f_{\Phi}(a)$, Eq.~(\ref{twoEquations}), namely the first term giving the dependence of the resonator frequency on the amplitude of motion, and the second term deriving from the feedback loop and present even for a linear resonator. The possibility of eliminating amplitude-phase conversion using a nonlinear resonator was demonstrated in Ref.~\cite{kenig12}, and is discussed for the phenomenological amplifier model in the next section.

\section{Phenomenological amplifier model}\label{phenomenological}

In this section we investigate the noise properties of an oscillator with the amplifier treated phenomenologically. We represent the amplifier by an instantaneous transfer function relating the drive on the resonator given by the output from the amplifier $q_{\text{out}}$ to the input to the amplifier $q_{\text{in}}$ given by the output from the resonator in the closed loop,\footnote{In the case of a mechanical resonator, we include the transduction from the displacement of the resonator into the electrical domain, and the electrical signal driving the displacement as part of the ``amplifier''.} so that
\begin{equation}
\label{d(t)}
d(t)=q_{\text{out}}=q_{s}\mathcal{A}(Gq_{\text{in}}(t)/q_{s}).
\end{equation}
The function $\mathcal A(y)$ is linear for small $y$, so that changing $G$ changes the linear gain of the amplifier. The function $\mathcal A(y)$ also describes the nonlinearity of the amplifier that develops for $q_{\text{in}}\sim q_{s}/G$, and saturates at the value 1 for large positive $y$, so that $q_{\text{out}}$ saturates at $q_{s}$.  As an example we use the amplifier function
\begin{equation}
\label{A(y)}
{\mathcal A}(y)=r\,\frac{1-e^{-2y}}{r+e^{-2y}},
\end{equation}
which gives a linear gain $2rG/(1+r)$ and saturation values $q_{s}$ for large positive $q_{\text{in}}$ and $-rq_{s}$ for large negative $q_{\text{in}}$. For $r=1$ the function reduces to ${\mathcal A}(y)=\tanh(y)$.

In addition, we include a phase shift (time delay) element which gives a phase shift in the fundamental of $\pi/2+\Delta$ so that $\Delta=0$ corresponds to positive feedback. In the following analysis, the phase shift is applied after the amplifier, although it could equally well be applied before.

\subsection{Gain function}

\begin{figure}[h]
\begin{center}
  \includegraphics[width=0.85\columnwidth]{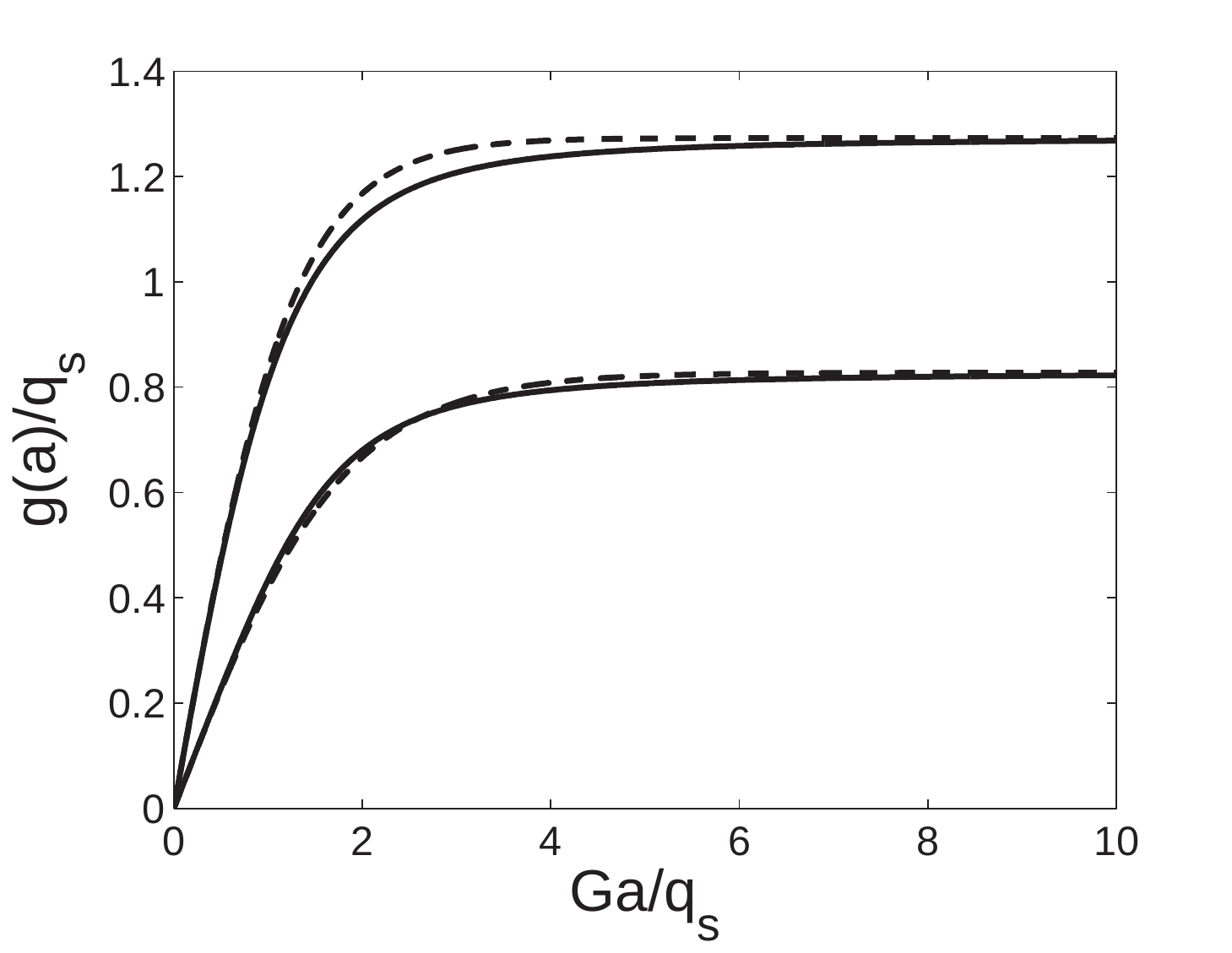}
  \caption{\label{fig: AmplifierCurves} Gain function $g(a)$ for the complex amplitude: solid curves - results from Eq.~(\ref{F(a)}) with the amplifier function $\mathcal A(y)$ given by Eq.~(\ref{A(y)}); dashed curves - approximate expression Eq.~(\ref{approximate g}). Results are shown for $r=1$ (upper curves) and $r=0.3$ (lower curves).}
  \end{center}
\end{figure}

To connect with the envelope treatment of the resonator in the closed loop, we need to determine the spectral components of the output signal of the amplifier for a periodic input $q_{\text{in}}(t)=a\cos(t+\Phi)$. The resulting drive on the resonator is
\begin{equation}
d(t)=\sum_nG_ne^{in(t+\Phi)},
\end{equation}
with the spectral components $G_{n}$ given by
\begin{equation}
  G_n =  \frac{q_{s}}{2\pi}\int^{\pi}_{-\pi}{\mathcal A}[Ga\cos x/q_{s}]\cos(nx)dx.
\end{equation}
Only the components $G_{1}=G_{-1}$ resonantly drive the resonator contributing to $D(T)$ in Eq.~(\ref{ampEqClean}), and the other components may be neglected. $G_{1}(a)$ then gives the effective gain function for a periodic signal at unit (scaled) frequency, so that the gain function in the equation for the complex amplitude (\ref{ampEqClean}) is
\begin{equation}
\label{F(a)}
g(a)=\frac{q_{s}}{\pi}\int^{\pi}_{-\pi}{\mathcal A}[Ga\cos x/q_{s}]\cos x\,dx.
\end{equation}
The gain function $g(a)$ is an odd function of $a$, and for the function $\mathcal A$ given in Eq.~(\ref{A(y)}) has the limits $g(a\to 0)=g_{l}a$ with the linear gain $g_{l}=2rG/(1+r)$, and saturation value $g(a\to\infty)=g_s=2q_{s}(1+r)/\pi$. For some purposes it is useful to have an approximate analytic expression for $g(a)$: as shown in Fig.~\ref{fig: AmplifierCurves} the function can be reasonably well approximated by
\begin{equation}
\label{approximate g}
g(a)\simeq g_s\tanh\left[\frac{g_l}{g_s}\,a\right].
\end{equation}

\subsection{Operating point}
\label{Sec: Operating point}

\begin{figure}[h]
\begin{center}
   \includegraphics[width=0.49\columnwidth]{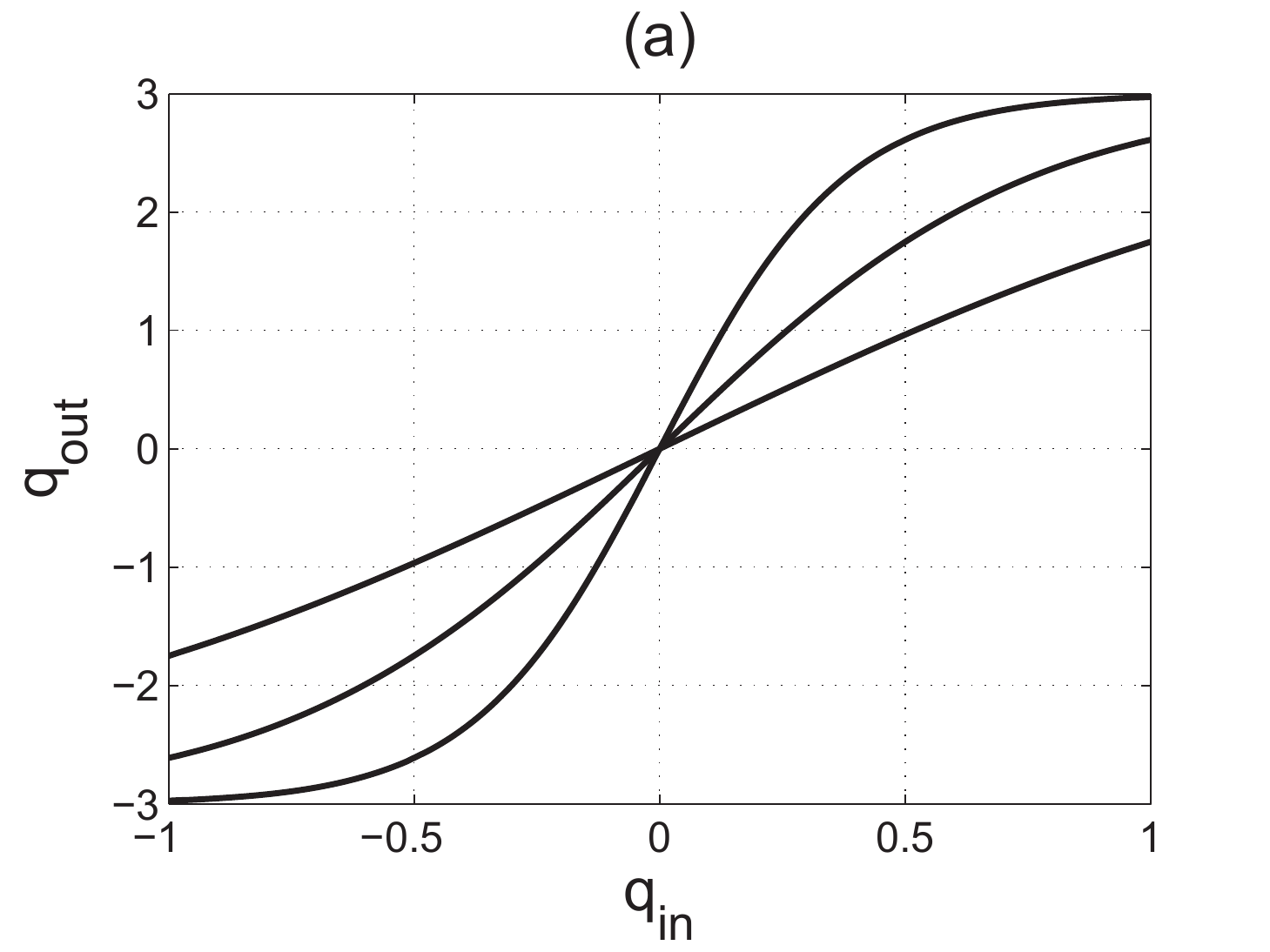}
   \includegraphics[width=0.49\columnwidth]{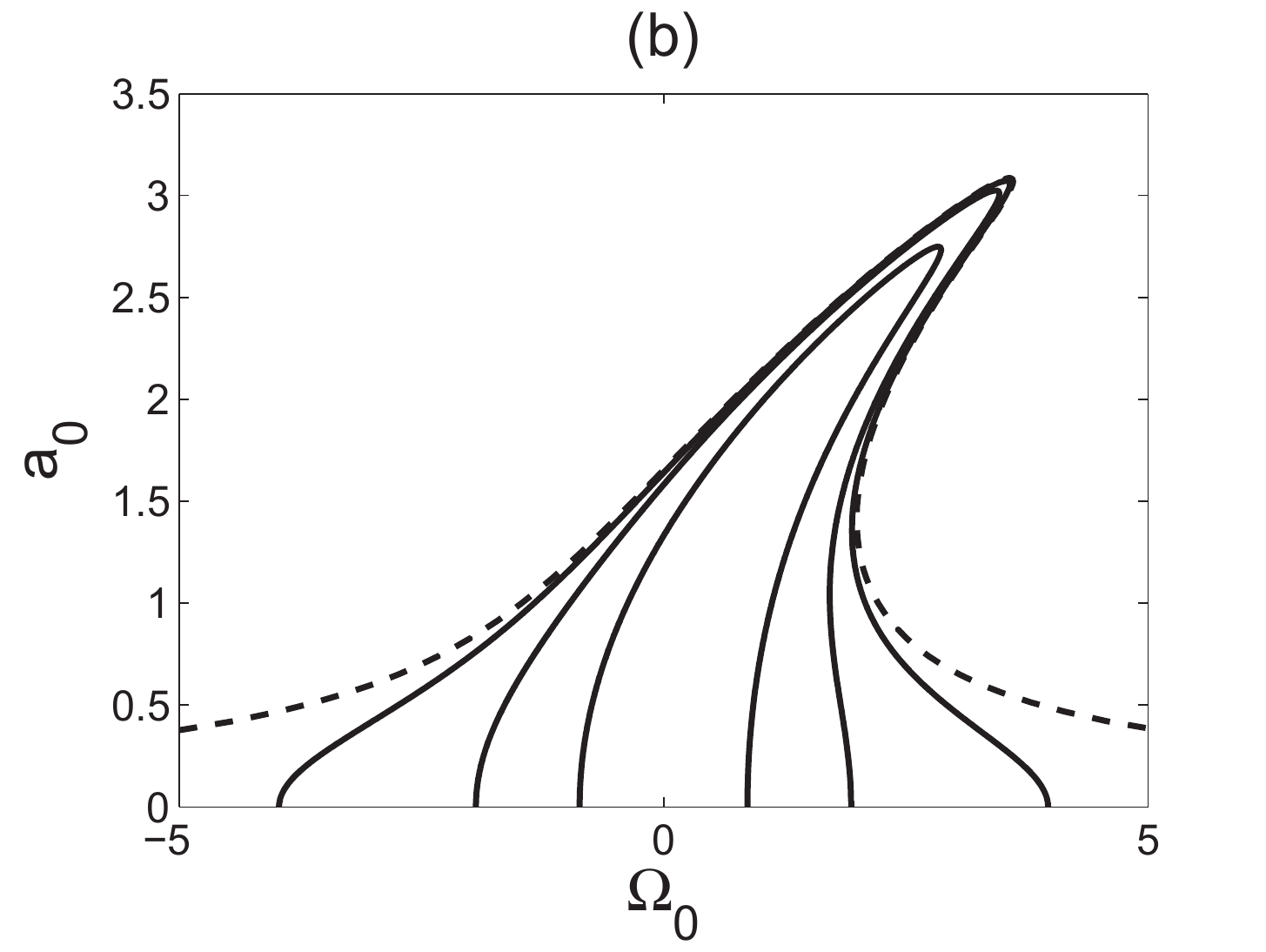}
   \includegraphics[width=0.49\columnwidth]{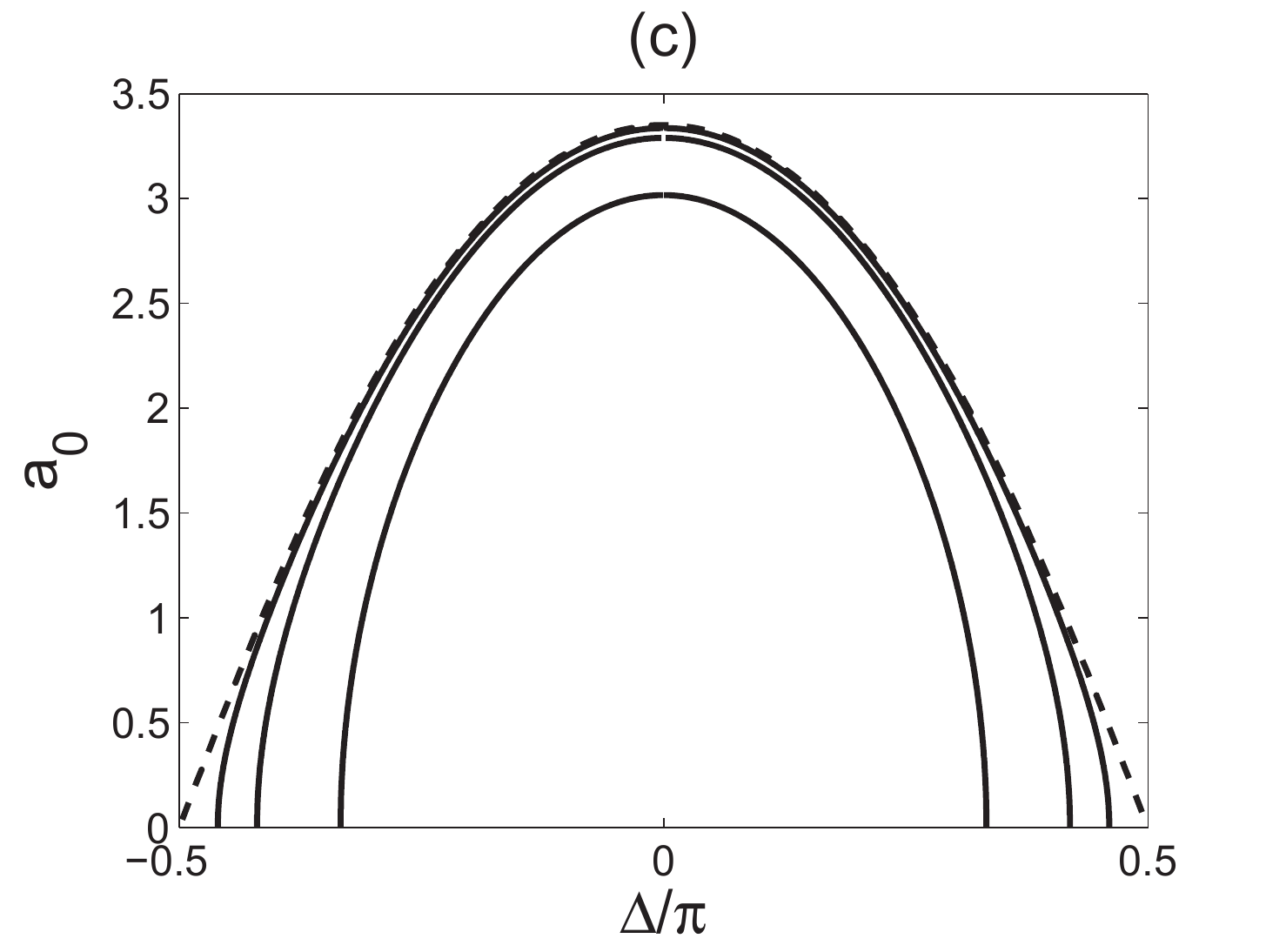}
   \includegraphics[width=0.49\columnwidth]{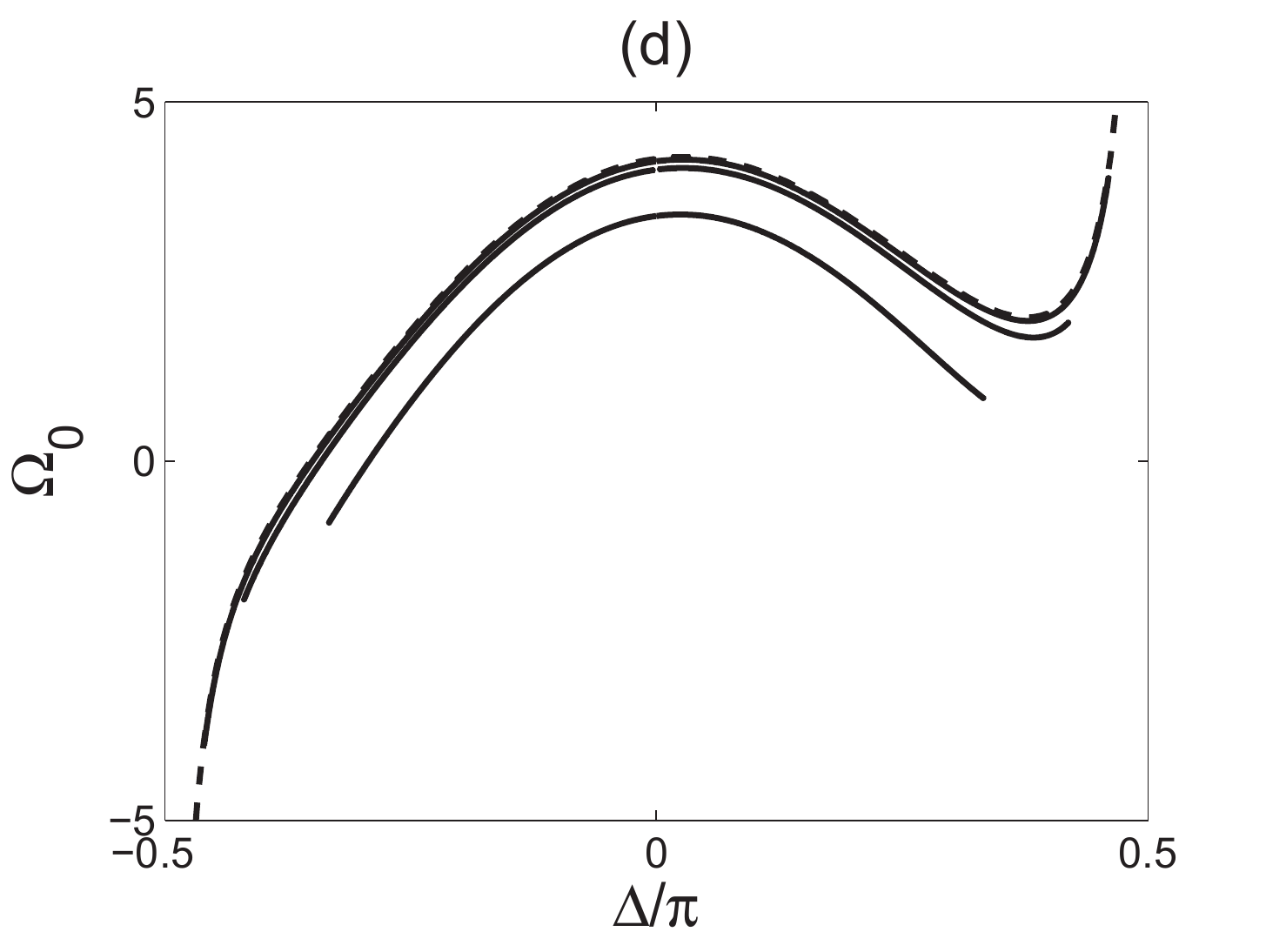}
   \caption{\label{phenomenological operating point} (a) Output vs. input curve for the amplifier given by Eqs.~(\ref{d(t)}) and (\ref{A(y)}); (b) oscillation amplitude $a_{0}$ vs.\ oscillation frequency $\Omega_{0}$ as the feedback phase shift is changed; (c) and (d) are the oscillation amplitude and frequency as a function of the feedback phase shift $\Delta$. The solid lines in (b), (c), and (d) are produced with Eqs.~(\ref{freqAmp}) and (\ref{F(a)}) for the gain values $G=2,4,$ and $8$ (values increase with increasing gain) and the dashed lines correspond to saturated gain limit and are given by Eqs.~(\ref{saturated}). Other parameters are $q_s=3$, $\alpha=1$, $\eta=0.1$, and $r=1$.}
      \end{center}
\end{figure}

Using Eqs.~(\ref{freqAmp}) with $g(a_{0})$ evaluated using Eq.~(\ref{F(a)}) gives the amplitude of oscillation $a_{0}$ and the oscillation frequency $\Omega_{0}$ as the phase shift $\Delta$ is varied. As an example, the oscillator response curve is shown in Fig.~\ref{phenomenological operating point} for $r=1,q_{s}=3$ and various values of the amplifier gain parameter, and using a value of the resonator nonlinear dissipation $\eta=0.1$. Note that the oscillations only occur over a limited range of feedback phases $\Delta$, with the range increasing with increasing $G>(1+r)/2r$ (given by unit linear gain $g_{l}=1$).

As the amplifier gain increases, the feedback approaches the constant level $g_s$, and in this limit there is an explicit expression for the oscillation amplitude and frequency as a function of the phase-shift
\begin{equation}
a_{0}=\eta^{-1/2}\bar a(g_{s}\eta^{1/2}\cos\Delta),\quad\Omega_0=f_{\Phi}(a_0),\label{saturated}
\end{equation}
with the function $\bar a$ given by
\begin{eqnarray}
\bar a(x)&=&\left(\frac{2}{9}\right)^{1/3}\left(\sqrt{3(27x^{2}+16)}+9x\right)^{1/3}\nonumber\\
&-&\left(\frac{32}{3}\right)^{1/3}\left(\sqrt{3(27x^{2}+16)}+9x\right)^{-1/3}.
\end{eqnarray}

\subsection{Noise models}

\subsubsection{Noise at the input to the amplifier}
As a first example of a noise source, we suppose there is an additive stationary noise at the input $q_{\text{in}}(t)\to q_{\text{in}}(t)+\xi_{s}(t)$. This noise is characterized by the single nonzero HPSD $Q_{s_0}(\omega)$. The output signal will also be noisy $ d(t)\to d(t)+\xi(t)$ with $\xi(t)$ the cyclostationary noise characterized by the HPSDs $Q_{l}(\omega)$. To relate the output noise to the input noise source we use the expression (\ref{spectrumHarmonics}).  For the amplifier function Eq.~(\ref{d(t)}) with input $q_{in}(t)=a_{0}\cos(t)$, we find the harmonic transfer functions of the amplifier $\bar H_{n}(\omega)=\bar H_{n}$ independent of $\omega$, with
\begin{eqnarray} \label{Hintegral}
  \bar H_{n} &=&   \frac{G}{2\pi}\int^{\pi}_{-\pi}{\mathcal A}'[Ga_{0}\cos x/q_{s}]\cos(nx)dx,
\end{eqnarray}
where ${\mathcal A}'(y)= d{\mathcal A(y)}/dy$. Including the effects of the phase shifter (which we assume does not add additional noise) gives the harmonic transfer function of the amplifier and phase shifter combination
\begin{equation}
\label{H_n phenomenological}
{H}_{n}=i^{n}e^{in\Delta} \bar H_{n}.
\end{equation}
These harmonic transfer functions are independent of the frequency.

\paragraph{Broadband noise}

\begin{figure}[htb]
\begin{center}
\includegraphics[width=0.9\columnwidth]{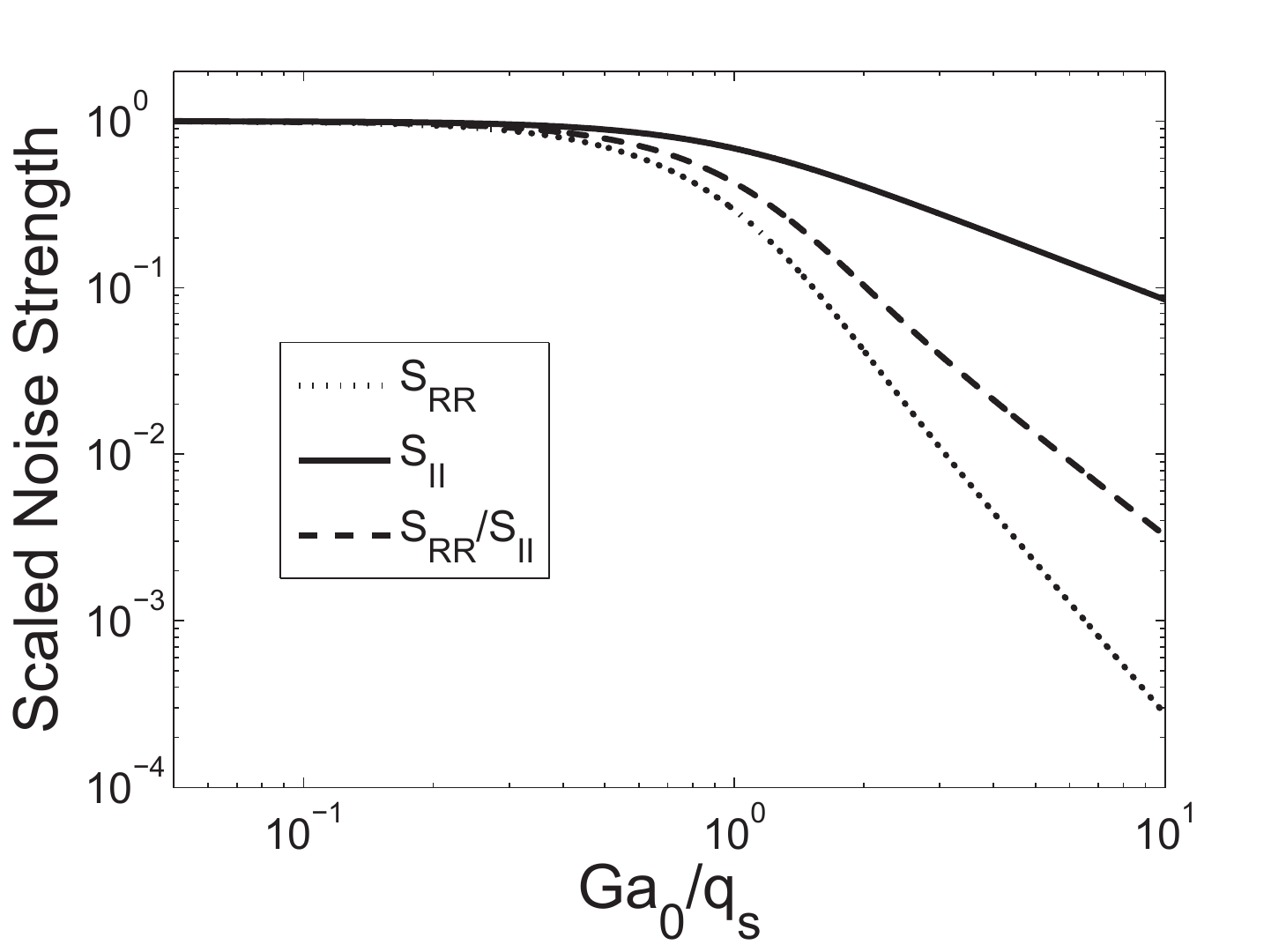}
\caption{Dependence of the slow noise strengths $S_{RR}, S_{II}$ on the amplifier nonlinearity according to Eqs.~(\ref{white}) for the amplifier function Eq.~(\ref {A(y)}) with $r=1$ and with white noise of strength $f_{0}$ at the amplifier input. The noise strengths are scaled by $2G^{2}f_{0}$ and are plotted as a function of $Ga_0/q_{s}$ for an input signal to the amplifier $a_0\cos t$.}
\label{fig:NoiseStrengths}
\end{center}
\end{figure}

Since all $ \bar H_{n} $ Eq.~(\ref{Hintegral}) are real, a convenient choice of the noise reference phase $\Phi_{N}$ that renders the cross correlation $S_{RI}^{s}$ in Eq.~(\ref {specs broadband}) zero is $\Phi_{N}=\Delta$, so that $\Xi_{R}$ gives the noise in the magnitude quadrature of the feedback drive and $\Xi_{I}$ the noise in the phase quadrature. Then Eqs.~(\ref{specs broadband}) can be written in the convenient form
\begin{eqnarray}\label{Sbroadcolored}
    S_{RR}&=&\sum_{n}Q_{s_{0}}(n)(\bar H_{n-1}+\bar H_{n+1})^{2},\\
    S_{II}&=&\sum_{n}Q_{s_{0}}(n)(\bar H_{n-1}-\bar H_{n+1})^{2}\nonumber,
\end{eqnarray}
where we have used the result $\bar H_{n}=\bar H_{-n},Q_{s0}(-n)=Q_{s0}(n)$. These equations describe the important effect the nonlinear amplifier has on the noise, converting the noise at the harmonics labeled by  $n$ to the carrier frequency.
Taking the high gain limit $(G\rightarrow\infty)$ in the Fourier coefficients of the amplifier derivative (\ref{Hintegral}) gives\footnote{Using the equation $\cos(nx)=T_n(\cos x)$ with $T_n(y)=\sum_{k=0}^{\lfloor n/2\rfloor} {n \choose 2k } (y^2-1)^ky^{n-2k}$, and $T_{2n}(0)=(-1)^{n},T_{2n+1}(0)=0$.}
\begin{equation}\label{highGainF}
    \bar{H}_{2n}\simeq(-1)^{n}\frac{q_s(1+r)}{\pi a_0},\quad \bar{H}_{2n+1}=0,
\end{equation}
which upon substitution into (\ref {Sbroadcolored}) gives $S_{RR}(\Omega)=0$, verifying for a saturated amplifier that the noise is entirely in the phase quadrature of the feedback.

For a white noise source,  $Q_{s_{0}}(\omega)=f_0$ independent of $\omega$, Eqs.~(\ref {Sbroadcolored}) simplify to the values
\begin{equation}\label{white}
    S_{RR}=2f_0(M_0+M_2),\quad S_{II}=2f_0(M_0-M_2),
\end{equation}
where
\begin {eqnarray}\label{Ml}
    M_l&=&\sum_{n}\bar H_n \bar H_{l-n}\nonumber\\
    &=&\frac{G^2}{2\pi}\int^{\pi}_{-\pi}\left\{{\cal A'}[Ga_{0}\cos x/q_s]\right\}^2\cos(lx)dx.
\end {eqnarray}
This gives
\begin{eqnarray}\label{}
   S_{RR}&=&2G^2f_{0}\ \frac{1}{\pi}\int^{\pi}_{-\pi}\left\{{\cal A'}[Ga_{0}\cos x/q_s]\right\}^2\cos^{2}x\,dx,\nonumber\\
    S_{II}&=&2G^2f_{0}\ \frac{1}{\pi}\int^{\pi}_{-\pi}\left\{{\cal A'}[Ga_{0}\cos x/q_s]\right\}^2\sin^{2}x\,dx.\nonumber\\
\end{eqnarray}
These functions are plotted in Fig.~\ref{fig:NoiseStrengths} for $r=1$. As the saturation level $Ga_{0}/q_{s}$ increases, the noise in the magnitude quadrature is suppressed, and the noise in the phase quadrature $S_{II}$ dominates. Note that for $r=1$, $M_2,\bar{H}_2\leq0$, so that the choice $\Phi_N=\Delta$ corresponds to $S_{RR}<S_{II}$.

\label{sec: phenomenological white noise}

For filtered white noise we take only the $n=\pm1$ terms in (\ref{Sbroadcolored}) and get
\begin{eqnarray}\label{filtered harmonics}
   \bar H_0+\bar H_2&=&\frac{G}{\pi}\int^{\pi}_{-\pi}{\mathcal A}'[Ga_{0}\cos x/q_{s}]\cos^2(x)dx=\left.\frac{dg}{da}\right|_{a=a_{0}}, \nonumber\\
   \bar H_0-\bar H_2&=&\frac{G}{\pi}\int^{\pi}_{-\pi}{\mathcal A}'[Ga_{0}\cos x/q_{s}]\sin^2(x)dx=\left.\frac{g}{a}\right|_{a=a_{0}},
\end{eqnarray}
(using integration by parts for the last equality)
so that the noise spectra are
\begin{equation}\label{SRR SII filtered}
    S_{RR}=2f_0\left.\left(\frac{dg}{da}\right)^2\right|_{a=a_{0}},\quad S_{II}=2f_0\left.\left(\frac{g}{a}\right)^2\right|_{a=a_{0}}.
\end{equation}
Since noise that is filtered around the oscillation frequency is equivalent to fluctuations in the signal itself, these feedback noise spectra can be directly obtained by linearizing the expression
\begin{equation}\label{}
    g(|A+\xi|)\frac{A+\xi}{|A+\xi|}e^{i\Delta}
\end{equation}
in the small complex noise $\xi$, as shown in Ref.~\cite{wiesenfeld13}.

\paragraph{1/f noise}
\label{sec:1/f}

\begin{figure}[tbh]
\begin{center}
\includegraphics[width=0.9\columnwidth]{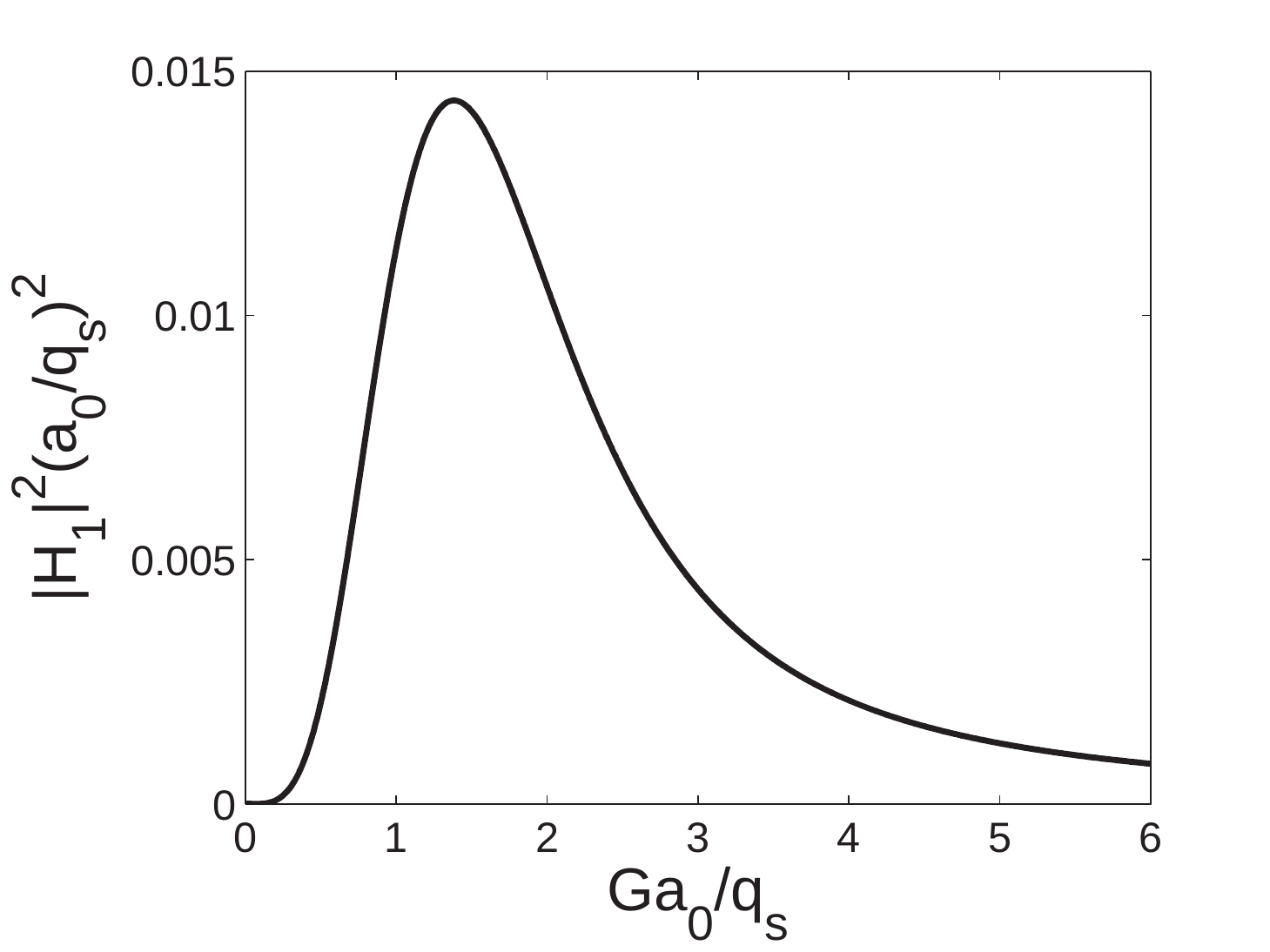}
\caption{Dependence of the slow noise strength $S_{RR}$ on the amplifier saturation nonlinearity for 1/f noise at the input to the amplifier specified by the function Eq.~(\ref {A(y)}) with $r=0.5$. The quantity $\bar H_{1}^{2}$ scaled by $(q_{s}/a_{0})^{2}$ is plotted as a function of $Ga_{0}/q_{s}$ for an input signal to the amplifier $a_{0}\cos t$. The full expression for $S_{RR}$ is given by Eq.~(\ref {S1/f}).}
\label{fig:H1}
\end{center}
\end{figure}

For 1/f noise we write $Q_{s_{0}}(\omega)=S_{1/f}(\omega)$, with, for example \cite{Demir02}
\begin{eqnarray}\label{}
S_{1/f}(\omega)&=&4f_0\int_{\omega_c}^\infty\frac{1}{x^{2}+\omega^2}dx\nonumber\\
 &=&\frac{2\pi f_0}{|\omega|}-\frac{4f_0\arctan(\omega_c/\omega)}{\omega},
\end{eqnarray}
cutting off the low frequency divergence below $\omega_c$ which is assumed small compared with $\varepsilon$, so that the 1/f spectrum extends well below the frequency corresponding to the resonator line width.
Equations (\ref{specs 1/f},\ref{H_n phenomenological}) now yield $\Phi_{N}=\Delta$ to eliminate $\Xi_{RI}^{s}$, and then Eqs.~(\ref{specs 1/f uncorrelated}) give
\begin{equation}\label{S1/f}
    S_{RR}(\Omega)=4 \bar H_1^2S_{1/f}(\varepsilon\Omega),\quad S_{II}(\Omega)=0.
\end{equation}
Note that amplifier noise in this case is represented by just one vector (the noise is in the magnitude quadrature of the feedback), and the spectrum of the noise is 1/f in the slow frequency. For the amplifier function Eq.~(\ref{A(y)}) the value of $\bar H_{1}$ is zero for $r=1$ giving an odd function $\mathcal A(y)$, since the up-conversion of the 1/f noise to the carrier frequency depends on the quadratic nonlinearity of the amplifier function. For $r\ne 1$, the contribution of the 1/f noise first increases with increasing gain or drive level as the quadratic nonlinearity becomes larger, and then decreases, since the amplifier saturation quenches the noise in the magnitude quadrature. An example for $r=0.5$ is shown in Fig.~\ref{fig:H1}.

As discussed in \S\ref{sec: 1/f general}, away from the carrier (larger $\Omega$) the slow noise may cross-over to a white-spectrum characterized by a different $\Phi_{N}$ and the strengths given in Eq.~(\ref {1/f noise tail}).

\subsubsection{Capacitor noise}

For additive 1/f noise at the amplifier input the slow noise vector is aligned with the feedback magnitude. In practical systems, there may be other 1/f sources which do not lead to such a simple result.  As an example of such a situation we consider 1/f noise in a capacitance in the feedback system. In a real system, this is likely to be an internal component of the amplifier, but purely for illustration we consider noise in a phase shifter implemented by an $RC$ filter after the amplifier with noise in the capacitor $C\rightarrow C(1+\xi_s)$. The feedback is taken as the voltage on the resistor.
The equation for the charge $q_{c}$ on the noisy capacitor is
\begin{equation}\label{current}
    \frac{dq_{c}}{dt}=-\frac{q_{c}-q_{c,in}}{\tau}+\frac{\xi_s(t)q_{c}}{\tau},
\end{equation}
where $q_{c,in}(t)$ is the input signal (in charge units) and $\tau$ is $RC$ scaled by the resonator frequency. Linearizing the solution to this equation in the noise strength leads to the harmonic transfer functions \cite{kenig1/f}
\begin{equation}\label{}
    H_1(\omega) =\frac{a_0\tau(1+\omega) e^{i(\Delta+\phi_c)}}{2\sqrt{\tau^2+1}\sqrt{\tau^2(1+\omega)^2+1}}, \quad H_{n\neq\pm1}(\omega)=0,
\end{equation}
with $\tan\phi_c=1/\tau$. Here $\Delta$ is the phase shift, including a contribution $\phi_{c}$ from the $RC$ filter (i.e.\ $\Delta=\phi_{c}$ if there are no other phase shifting elements).
For broadband noise we substitute this in Eqs.~(\ref{specs broadband uncorrelated}) with $Q_{s_{0}}(\omega)=f_0$, which gives
\begin{equation}\label{}
    S_{RR}=6f_0|H_1(0)|^2,\quad S_{II}=2f_0|H_1(0)|^2,
\end{equation}
whereas for 1/f noise we get
\begin{equation}\label{}
    S_{RR}(\Omega)=4|H_1(0)|^2S_{1/f}(\varepsilon\Omega),\quad S_{II}(\Omega)=0,
\end{equation}
where in both cases $\Phi_N=\Delta+\phi_c-\pi/2$. Again, we witness the result that broadband noise is expressed as a ball in the complex amplitude phase space, whereas 1/f noise is just a line. The direction of the noise ball (broadband) or line (1/f) is now aligned at an angle $\phi_{c}-\pi/2$ to the direction of the feedback.

\subsection{Oscillator phase noise}

We now present results for the oscillator phase noise, focusing in particular on special operating points of the oscillator where the detrimental effects of the amplifier and resonator noise are reduced or even eliminated using the nonlinear behavior of the resonator. The procedure is a follows. For a particular noise source, the slow noise forms a ball in the resonator phase space with axes making an angle $\Phi_{N}$ to the complex amplitude, and with uncorrelated noises with spectra $S_{RR}(\Omega)$ and $S_{II}(\Omega)$ along and perpendicular to the direction defined by $\Phi_{N}$. The quantities $\Phi_{N},S_{RR},S_{II}$ were calculated for various noise sources in the previous section. The resulting phase noise is then given by the projection of these two independent noises along the phase sensitivity vector of the resonator, given by Eq.~(\ref{zeroMode}) with the amplitude $a_{0}$ given by the operating point \S\ref {Sec: Operating point}. We focus in particular on the dependence on the feedback phase induced by $P_{\text{eff}}^{2}(\Delta)$, Eq.~(\ref {P_eff}). In special cases, one of the strengths $S_{RR},S_{II}$ may be zero, so that the noise acts along a line rather than filling a ball. In this case it may be possible to \emph{eliminate} the effects of a particular noise source by tuning $\Delta$ to make $P_{\text{eff}}=0$.

\subsubsection{White feedback noise}

\begin{figure}[htb]
  \includegraphics[width=0.95\columnwidth]{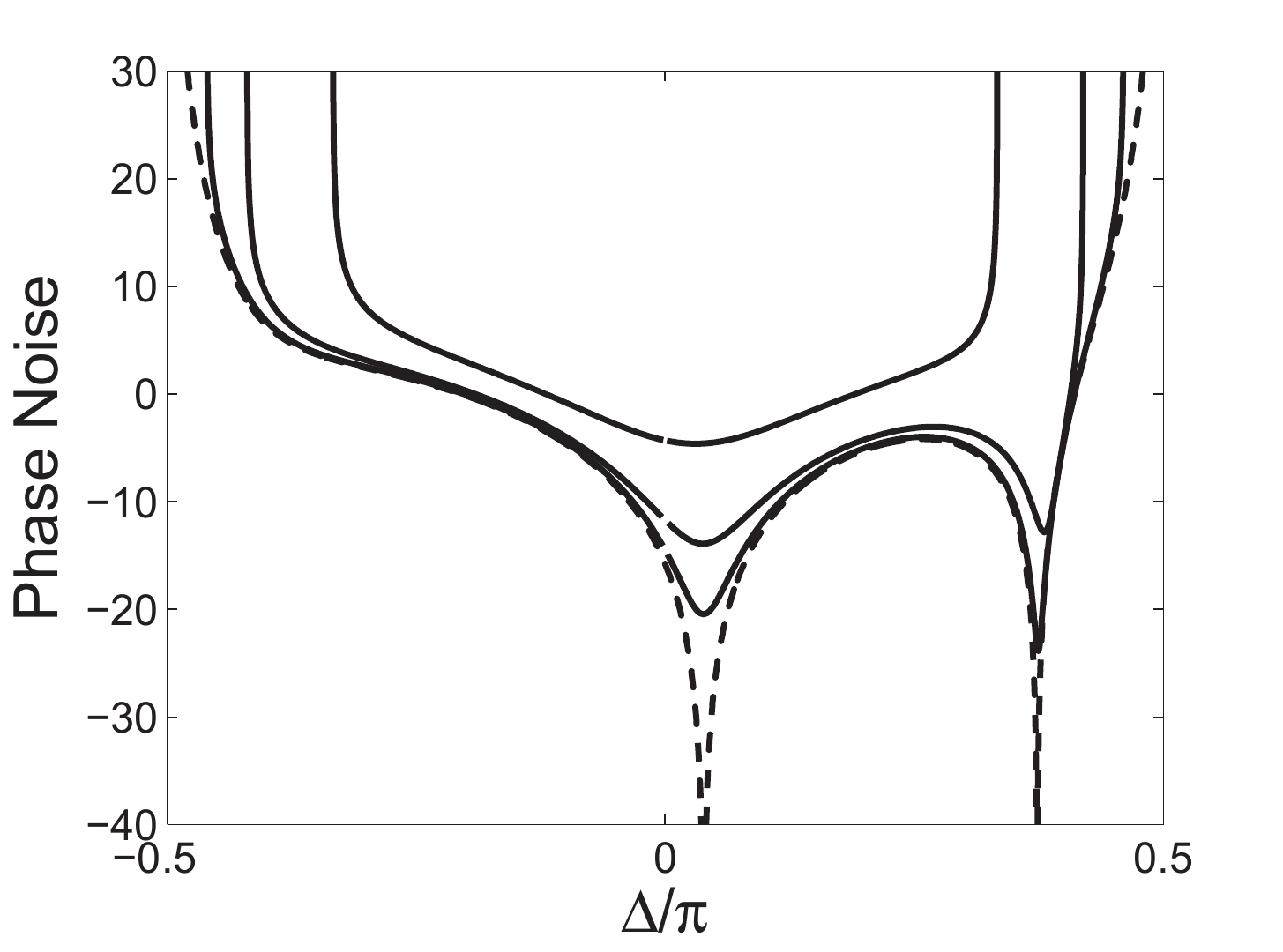}
  \caption{\label{phaseNoiseFigure} Dependence of the phase noise on the feedback phase for a white noise source at the amplifier input. The quantity  $10\log_{10}P_{\textmd{eff}}^2$, Eq.~(\ref{wt}), is plotted for the same parameters as in Fig.~\ref{phenomenological operating point}: solid line -- gain values $G=2,4,$ and $8$ (phase noise curves decreasing with increasing gain); dashed line -- high gain limit where the feedback level is constant.}
\end{figure}

To calculate the oscillator phase noise due to a white noise source at the amplifier input we use the results from Section \ref{sec: phenomenological white noise} to give the phase variance Eq.~(\ref{variance white noise}) $V(\tau)=4\varepsilon f_{0}M_{0}P_{\textmd{eff}}^2\,\tau$ and the phase noise spectrum away from the carrier Eq.~(\ref {phase noise spectrum})
$\bar S=4 f_0M_{0}P_{\textmd{eff}}^2/\Omega^{2}$ with
\begin{equation}\label{wt}
    P_{\textmd{eff}}^2=\frac{1}{2}\left[P_R^2\left(1+\frac{M_2}{M_0}\right)+P_I^2 \left(1-\frac{M_2}{M_0} \right)\right].
\end{equation}
The quantity $P_{\textmd{eff}}^2$ giving the $\Delta$ dependence of the oscillator phase noise is plotted in Fig.~\ref{phaseNoiseFigure} for the symmetric amplifier, $r=1$, and various values of the amplifier gain. In the high gain limit the $M_{l}$ Eq.~(\ref{Ml}) can be evaluated explicitly\footnote{Note that for a direct evaluation of the sum in Eq.~(\ref{specs broadband uncorrelated}) with $Q_{s_{0}}\to\text{const.}$, each $H_{n}$ is independent of $G$ for $G\to\infty$ but the sum then diverges.}
\begin{equation}
  M_0 = -M_{2}= \frac{4Gq_s}{3\pi a_0},
\end{equation}
and substituting these into (\ref{wt}) gives
\begin{equation}\label{sat}
    P_{\textmd{eff}}^2=P_I^2=\frac{ \pi^{2}}{16q_{s}^{2}}\left[\frac{d\Omega_0}{d\Delta}\right]^2,
\end{equation}
using Eq.~(\ref{PI}) and $g(a_{0}\to\infty)=4q_{s}/\pi$.
Since the amplifier is saturated in this limit, there is only noise in the phase of the feedback, and the condition $d\Omega_{0}/d\Delta=0$ defines two operational points for which all the feedback noise is eliminated \cite{kenig12,YurkePra}, as shown in Fig.~\ref{phaseNoiseFigure}. This condition is supported by phase noise measurements of a NEMS oscillator with a saturated amplifier \cite{Villanueva13}. Fig.~\ref{phaseNoiseFigure} also shows that significant noise reduction is achieved for the unsaturated amplifier in the vicinity of these points.
The results for filtered white noise at the amplifier input, given by Eqs.~(\ref{SRR SII filtered}), show similar trends, and are plotted in Ref.~\cite{wiesenfeld13}.

\subsubsection{1/f feedback noise}
\begin{figure}
  \includegraphics[width=0.95\columnwidth]{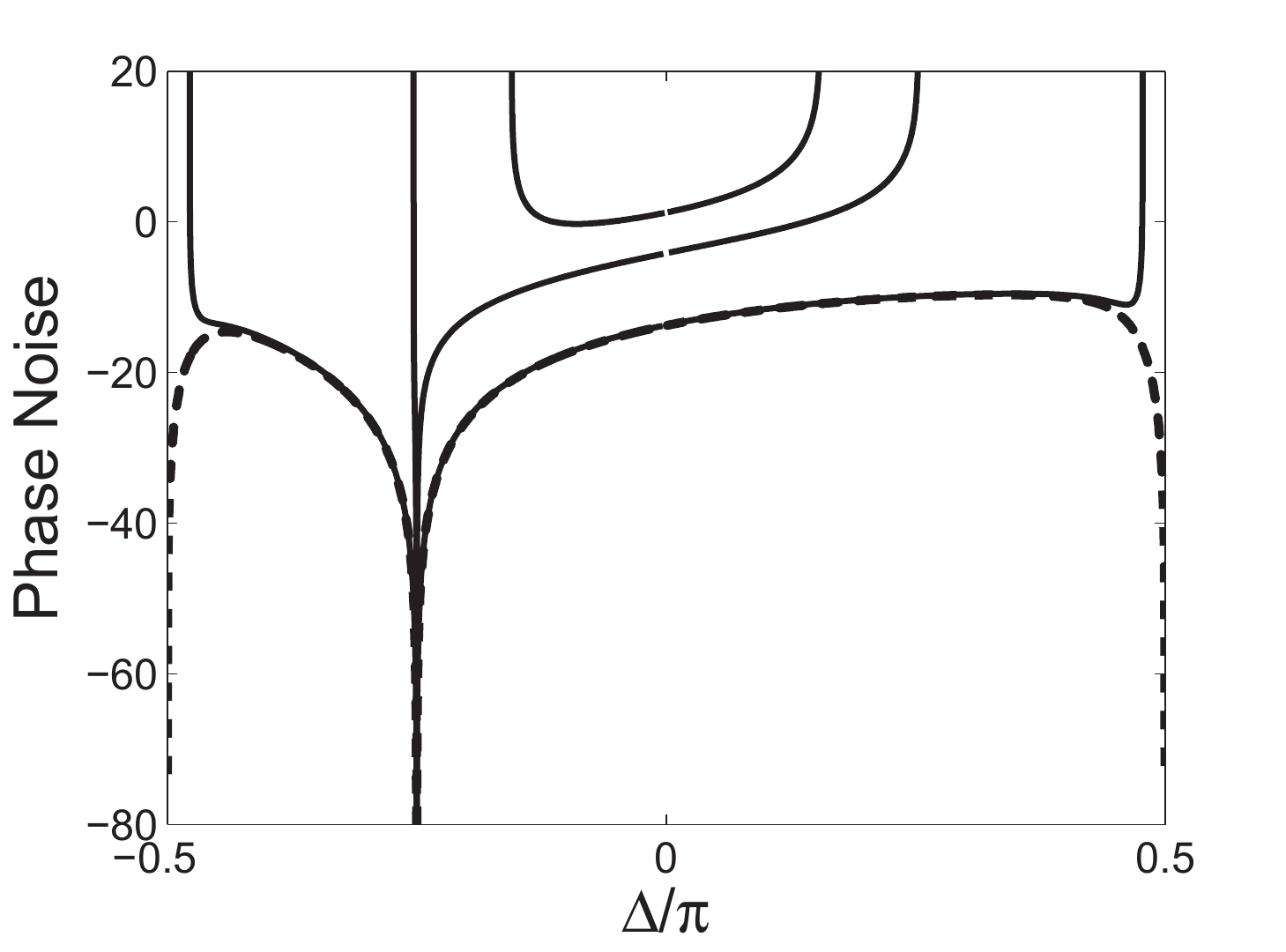}\\
  \caption{\label{zeroPhaseElimination} Dependence of the phase noise on the feedback phase for a 1/f noise source at the amplifier input. The quantity $10\log_{10}P_{R}^2$ is plotted: solid curves -- gain levels $G=(0.8,1.01,10)\cdot G_c $ with $G_c\simeq2.12$ (phase noise decreasing with increasing gain); dashed curve -- the high gain limit where the feedback level is constant. Other parameter values are $r=0.5$, $q_s=3$, $\alpha=1$, and $\eta=3$. The phase shift value which gives zero noise is $\Delta_R=-\pi/4$.}
\end{figure}

In Section \ref{sec:1/f} we showed that the leading order effect of 1/f noise at the input of the phenomenological amplifier is 1/f slow noise acting on the complex amplitude that is purely in the magnitude quadrature of the feedback $\Phi_{N}=\Delta,\Xi_{I}=0$.
Thus if the phase sensitivity $P_{R}$ can be tuned to zero, the effect of the 1/f noise on the oscillator phase noise is eliminated. For the model we are considering, it turns out from Eqs.~(\ref{twoEquations},\ref{noiseVectors},\ref{phaseSensitivity},\ref{zeroMode}) that $P_{R}$ is zero for the phase shift value $\Delta=\Delta_R=-\arctan\left(3\alpha/\eta\right)$, independent of the amplifier parameters. Combining this result with the requirement for positive oscillation amplitude $a>0$ yields the condition on the amplifier gain $G>G_c=(1+r)\sqrt{1+9\alpha^{2}/\eta^2}/2r$.
If the oscillator parameters satisfy this condition, 1/f noise at the amplifier input can be eliminated by tuning to the special value $\Delta_{R}$. The full behavior of $P_{R}^{2}$ as a function of the feedback phase and for various amplifier gain parameters is shown in Fig.~\ref{zeroPhaseElimination}. Note that in contrast to Fig.~\ref{phaseNoiseFigure} for white noise, the phase noise resulting from the 1/f noise source can be completely eliminated even using an unsaturated amplifier.

In the more general case for which $\Phi_N\neq\Delta$, as in our example of capacitor noise, the ability to eliminate 1/f noise depends on the parameters characterizing the amplifier. This was studied in Ref.~\cite{kenig1/f}.

\subsubsection{Linear amplifier}
\label{sec:linear amp}
If the resonator has nonlinear damping, the closed loop oscillator can be constructed using a linear amplifier $g(a)=Ga$, with the resonator providing the saturating nonlinearity. The HSPDs of such an amplifier are $\bar H_{0}=G,\bar H_{n\ne 0}=0$, and Eq.~(\ref{twoEquations}) gives the expression for the oscillation amplitude in this limit $a_0^2=4(G\cos\Delta-1)/\eta$.  For white noise at the amplifier input the phase noise can be calculated explicitly. For example the phase variance Eq.~(\ref{variance white noise}) is
\begin{equation}
V(\tau)=\varepsilon  f_{0}\left[\frac{9\alpha^{2}+\eta^{2}}{8\eta}\frac{G^{2}}{G\cos\Delta-1}\right]\tau,
\end{equation}
with a corresponding result for the spectrum $\bar S$, Eq.~(\ref{phase noise spectrum}).
Although this expression cannot be tuned to zero, it is minimized (for fixed $\alpha$) for $\Delta=0$, $G=2$, and $\eta=3\alpha$, when the factor in the braces becomes $3\alpha$.
1/f noise at the amplifier input does not contribute to the oscillator phase noise, since there is no up conversion of 1/f noise by the linear amplifier ($\bar H_{1}=0$).

\subsubsection{Resonator noise}
\begin{figure}[h]
\begin{center}
  \includegraphics[width=0.9\columnwidth]{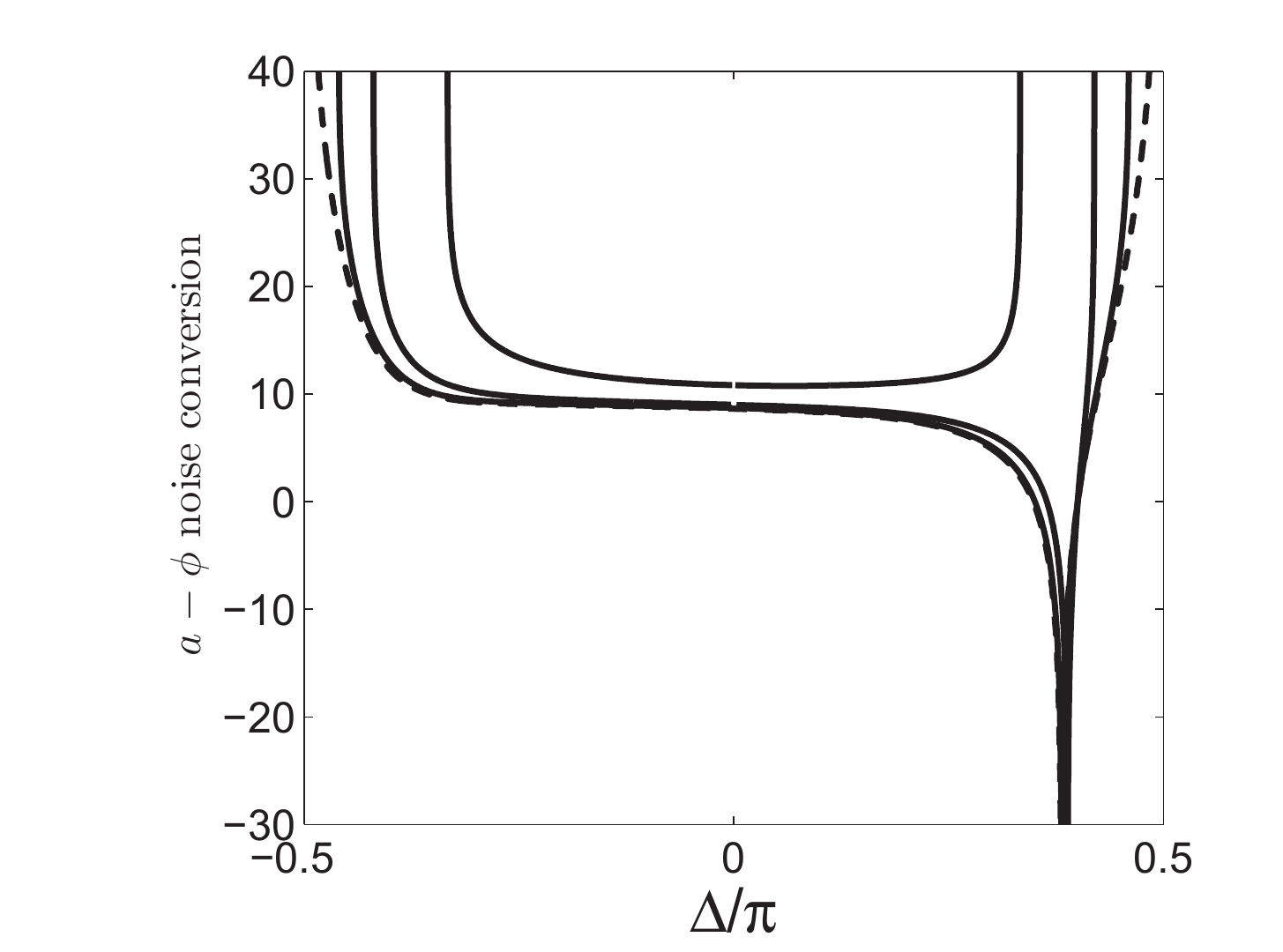}
  \caption{\label{fig: amplitudePhase} Amplitude-phase noise conversion factor as a function of the feedback phase $\Delta$. The quantity $10\log_{10}[(f'_{\phi}(a_{0})/f'_{a}(a_{0}))^{2}]$ is plotted for the same parameters as in Figs.~\ref{phenomenological operating point},\ref{phaseNoiseFigure}: solid curves -- gain values $G=2,4,8$ (amplitude-phase conversion decreasing with increasing gain); dashed curve -- the high gain limit where the feedback level is constant. Other parameter values are $\alpha=1,\eta=0.1,q_s=3,r=1$. Note that the amplitude-phase noise conversion is eliminated for $\Delta/\pi\simeq0.4$.}
  \end{center}
\end{figure}

As discussed in \S\ref{Incomplete}, the oscillator phase noise induced by noise forces in the magnitude quadrature of the resonator motion (amplitude-phase conversion) can be eliminated where $f'_\Phi(a_{0})=0$. Since the resonator noise sources are independent of the feedback phase, the full $\Delta$ dependence of this contribution is given by the square of the first (amplitude) component of $\mathbf{v}_{\perp}$ in Eq.~(\ref{zeroMode}), i.e.\ by $[f'_{\phi}(a_{0})/f'_{a}(a_{0})]^{2}$. This is plotted in Fig.~\ref{fig: amplitudePhase} for the same parameters as in Figs.~\ref{phenomenological operating point},\ref{phaseNoiseFigure}. An interesting feature is that the point of strong noise suppression ($\Delta/\pi\simeq0.4$) is close to the right dip in the phase noise in Fig.~\ref{phaseNoiseFigure}. These two points actually approach each other for $\eta=0$ and a saturated amplifier with a constant feedback level $g_s$ in the limit $g_s\rightarrow\infty$, and this can be exploited to suppress both amplifier noise and the magnitude quadrature of the resonator noise as shown theoretically and experimentally in \cite{kenig12,Villanueva13}. Since noise in the damping coefficient (originating from a fluctuating resistor in an electronic resonator circuit, for example) is given by Eqs.~(\ref{parameterNoise}) with $S_{RR}$, $S_{II}$ interchanged, 1/f noise in this coefficient is eliminated where $f'_\Phi(a_{0})=0$. As discussed before, the phase quadrature of the resonator noise acts directly on the oscillator phase noise, and cannot be quenched by tuning $\Delta$, although its effect is reduced by going to large oscillation amplitudes.

\subsubsection{Linear resonator}
\begin{figure}[h]
\begin{center}
  \includegraphics[width=0.6\columnwidth]{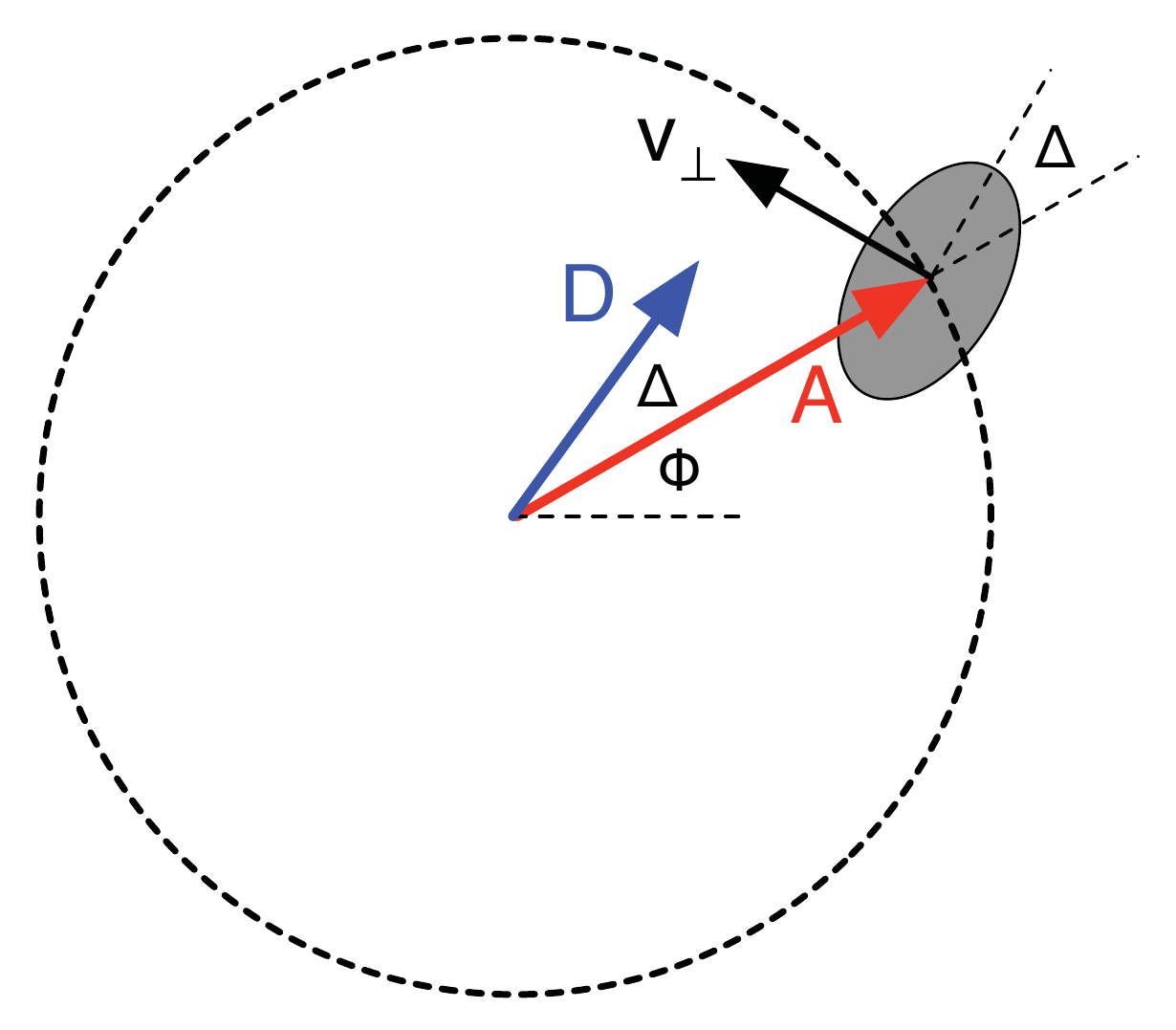}
  \caption{\label{fig: NoiseEllipseLinear} Noise vectors in the complex amplitude phase space for a linear resonator and a noise source at the input to the amplifier. The red arrow is the resonator amplitude $A$, the blue arrow is the feedback drive $D$ with a phase shift $\Delta$ from the resonator signal, and the black arrow is the phase sensitivity vector $\mathbf{v}_{\perp}$. The shaded ellipse shows the alignment of the noise ball, with axes parallel and perpendicular to the feedback direction. Note that for a linear resonator $\mathbf{v}_{\perp}$ is perpendicular to the feedback vector and the noise ball axis corresponding to $\Xi_{R}$: this is not in general true for a nonlinear resonator.}
  \end{center}
\end{figure}

It is interesting to compare the results of the previous sections with those for a linear resonator, given by setting $\alpha=\eta=0$ in Eqs.~(\ref{ampEq},\ref{twoEquations}). The phase sensitivity vector Eqs.~(\ref{twoEquations},\ref{zeroMode}) becomes
\begin{equation}
\mathbf{v}_{\perp}=(-a_{0}^{-1}\tan\Delta,1).
\end{equation}
This vector is perpendicular to the feedback vector, as shown in Fig.~\ref{fig: NoiseEllipseLinear}, i.e.\  if we choose $\Phi_{N}=\Delta$ then $\mathbf{v}_{\perp}\cdot\mathbf{v}_{R}=0$. This means that feedback noise in the magnitude quadrature does not contribute to the oscillator phase noise, leaving only the contribution from noise in the phase quadrature. Using the noise vectors Eq.~(\ref{noiseVectors}) with $\Phi_{N}=\Delta$ then gives a simple explicit expression for the $\Delta$ dependence of the oscillator phase noise proportional to $P_{I}^{2}S_{II}=\tfrac{1}{4}a_{0}^{-2}\sec^{2}\Delta\,S_{II}$ ($S_{II}$ may depend on $\Delta$ due to the amplitude dependent mixing behavior of the amplifier).

Thermodynamic resonator noise has a similar dependence of the phase noise on $\Delta$ for the linear resonator: this noise source is isotropic, and any choice of $\Phi_{N}$ gives uncorrelated noise in the two quadratures with $S_{RR},S_{II}$ equal. This gives the oscillator phase noise proportional to the same expression $\tfrac{1}{4}a_{0}^{-2}\sec^{2}\Delta\,S_{II}$, now with $S_{II}=\tfrac{1}{2}S_{\text{tot}}$ independent of $\Delta$. These expressions for the oscillator phase noise show the familiar $a_{0}^{-2}$ reduction as the resonator amplitude grows. In addition there is a worsening of the phase noise proportional to $\sec^{2}\Delta$ if the feedback phase is shifted away from the value $\Delta=0$ giving maximum oscillation amplitude. In either case, the phase noise only has a smooth dependence on $\Delta$ showing that resonator nonlinearity is crucial to the dramatic noise reduction near special operating points seen in Fig.~\ref{phaseNoiseFigure}.

Note that the phase sensitivity vector $\mathbf{v}_{\perp}$ does \emph{not} lie along the limit cycle in Fig.~\ref{fig: NoiseEllipseLinear}, so that there is amplitude-phase noise conversion even for the linear resonator. Also note that sending the amplifier to the saturated limit does not open the possibility of complete noise elimination for a linear resonator.

\section{Realistic amplifier models}\label{Sec: simulation}

In this section we extend the results of Section \ref{phenomenological} to realistic descriptions of the amplifier.
If the noise sources of the amplifier can be well modeled in terms of stationary noise added at the input of the amplifier, the formalism of Section \ref{phenomenological} is readily generalized. This may often be a good approximation, since noise from early stages of a compound amplifier, will be amplified most, and may dominate noise from later stages. The output noise is then given by Eqs.~(\ref{specs broadband uncorrelated},\ref{specs 1/f uncorrelated}) with the $H_n$ given by the harmonic transfer functions for the amplifier with an input $a_{0}\cos t$. These may readily be calculated given a model of the amplifier, perhaps using a circuit simulator package.

More generally, there may be additional noise sources from the internal components of the amplifier.  Rather than calculating the various harmonic transfer functions $H_n$ and combining them to form the output noise, we develop an approach to directly find the slow noise terms using either circuit simulator calculations or experimental measurements of the amplifier in an appropriate open loop configuration.

In this approach the amplifier is driven with a signal $a\cos\omega_{0}t$ with $a$ the amplitude for a particular operating point of the closed loop oscillator. The output of the amplifier is then mixed with a copy of the input signal shifted by a phase $\Phi_{M}$, and the result is fed into a low pass filter which transmits frequencies $\omega<\omega_{b}$ with $\omega_{b}\ll\omega_{0}$. The noise in the output of this combination is then the desired amplitude equation noise $\Xi$.

We now show how this setup reproduces the amplitude equation noise. The generalization of Eq.~(\ref{spectrumHarmonics}) for the case of cyclostationary input noise is \cite{Roychowdhury98}
\begin{equation}\label{GeneralspectrumHarmonics}
    Q_{o_l}(\omega)=\sum_{n,k}H_n(-\omega-n)Q_{i_k}(\omega+n)H_{l-n-k}(\omega+n+k).
\end{equation}
Suppose the noise at the output of the amplifier is characterized by the HPSDs $Q_{a_l}(\omega)$. After mixing this signal with $\cos(t+\Phi_{M})$, the HPSDs of the noise at the output of the mixer are
\begin{equation}\label{}
   Q_{m_l}(\omega;\Phi_{M})=\sum_{k,n}H^{(m)}_nH^{(m)}_{l-n-k}Q_{a_k}(\omega+n)e^{i(l-k)\Phi_{M}},
\end{equation}
with $H^{(m)}_{n}$ here the harmonic transfer functions of the mixer which are $H^{(m)}_{n=\pm1}=1/2$, and $H^{(m)}_n=0$ otherwise \cite{Roychowdhury98}. This gives
\begin{eqnarray}\label{}
   Q_{m_l}(\omega;\Phi_{M})=\tfrac{1}{4}\big[Q_{a_l}(\omega-1)&+&Q_{a_l}(\omega+1)\\
   +Q_{a_{(l+2)}}(\omega-1)e^{-2i\Phi_{M}}&+&Q_{a_{(l-2)}}(\omega+1)e^{2i\Phi_{M}} \big].\nonumber
\end{eqnarray}
Now we perform the low frequency filtering of this signal to extract just the stationary $l=0$ component, which is
\begin{eqnarray}\label{}
   Q_{m_0}(\omega;\Phi_{M})=\tfrac{1}{4}\big[Q_{a_0}(\omega-1)&+&Q_{a_0}(\omega+1)\\
   +Q_{a_2}(\omega-1)e^{-2i\Phi_{M}}&+&Q_{a_{-2}}(\omega+1)e^{2i\Phi_{M}}  \big].\nonumber
\end{eqnarray}
The slow noise spectra, Eqs.~(\ref{spec real imaginary and cross}) in the Appendix, can now be calculated by
\begin{eqnarray}
 S_{RR}(\Omega) &=& 4 Q_{m_0}(\varepsilon\Omega;\Phi_{N}+\tfrac{\pi}{2}), \\
 S_{II}(\Omega) &=& 4 Q_{m_0}(\varepsilon\Omega;\Phi_{N}+\pi), \nonumber\\
 S_{RI}^{s}(\Omega) &=& 4[Q_{m_0}(\varepsilon\Omega;\Phi_{N}+\tfrac{3\pi}{4})-Q_{m_0}(\varepsilon\Omega;\Phi_{N}+\tfrac{\pi}{4})].\nonumber
\end{eqnarray}
The phase noise of the closed loop oscillator is given by substituting these quantities into (\ref{phaseVariance}) or (\ref{phase noise spectrum}).
\appendices
\numberwithin{equation}{section}
\section{Deriving the phase evolution equation}
\label{Appendix: phase equation}

In this appendix we derive the phase evolution equation (\ref{stochastic phase}) and the expression (\ref{phaseSensitivity}) for the noise sensitivity vector.   We use the method of secular perturbation theory, essentially following the approach of Demir et al.\ \cite{DemirMehrotra00}, but in the simpler context of our discussion where the limit cycle solution can be determined as the fixed point of dynamical equations. The Floquet stability analysis of the limit cycle then reduces to the simpler discussion of the stability of a fixed point.

We could present the method using the two component vector $\mathbf{x}\equiv(a,\Phi)$ notation Eq.~(\ref{X EOM}). Instead, we will formulate the argument in the Cartesian space $\mathbf{X}\equiv(X,Y)\equiv(a\cos\Phi,a\sin\Phi)$, since this more closely follows Ref.~\cite{DemirMehrotra00}.  In these coordinates, the limit cycle representing the oscillating solution in the absence of noise is a point rotating at rate $\Omega_{0}$ around a circle of radius $a_{0}$ given by the operating point $f_{a}(a_{0})=0,f_{\Phi}(a_{0})=\Omega_{0}$, see \S\ref{Sec: Operating point}.

To simplify the analysis, we go to a coordinate system rotating at the rate $\Omega_{0}$.
In this rotating frame the equation of motion (\ref{ampEqClean}) of the complex amplitude  takes the form
\begin{equation}
\label{Eqn: stochastic X}
\frac{d\mathbf{X}}{dT}=\mathbf{F}(a)+\boldsymbol\Xi,
\end{equation}
with $\mathbf{F}$ the deterministic terms and $\boldsymbol\Xi$ the noise terms. Because of the phase invariance of the amplitude equation (\ref{ampEqClean}), in the rotating frame $\mathbf{F}$ has no explicit time dependence and depends only on the magnitude $a=|\mathbf{X}|$, and the statistics of the noise $\boldsymbol\Xi$ are stationary. The limit cycle is given by the fixed point $\mathbf{X}_{0}(\Phi_{0})=(a_{0}\cos\Phi_{0},a_{0}\sin\Phi_{0})$ determined by $\mathbf{F}(a_{0})=0$, and depending on an arbitrary phase $\Phi_{0}$.

Now consider the effect of the noise $\boldsymbol\Xi$, assumed small. Since there is no restoring force on the phase $\Phi$, even small noise may generate a large phase change over long times. We therefore write the solution to the stochastic equation as
\begin{equation}
\mathbf{X}(T)=\mathbf{X}_{0}(\Phi(T))+\mathbf{X}_{1}(T),
\end{equation}
where $\mathbf{X}_{1}(T)$ is a small correction and the evolution of the phase is slow (i.e.\ large changes in $\Phi$ take a long time $T$), both related to the small strength of the noise. Substituting into the stochastic equation of motion (\ref{Eqn: stochastic X}) leads to the equation linearized for the small correction $\mathbf{X}_{1}$
\begin{equation}
\label{Eqn: stochastic X1}
\frac{d\mathbf{X_{1}}}{dT}-\mathbf{J}\cdot\mathbf{X_{1}}=-\left.\frac{d\mathbf{X}_{0}}{d\Phi}\right|_{\mathbf{X}=\mathbf{X_{0}}}\frac{d\Phi}{dT}+\boldsymbol\Xi,
\end{equation}
with $\mathbf{J}$ the Jacobian of the linear stability analysis.
\begin{equation}
J_{ij}=\left.\frac{\partial{F_{i}}}{\partial{X_{j}}}\right|_{\mathbf{X}=\mathbf{X_{0}}}.
\end{equation}
There are two stability eigenvectors of the Jacobian. There is a zero eigenvalue eigenvector corresponding to the arbitrary phase of the fixed point solution. This vector can be written (with a choice of normalization corresponding to a unit phase change)
\begin{equation}
\label{Eqn: e0}
\mathbf{e}_{0}=\left.\frac{d\mathbf{X}_{0}}{d\Phi}\right|_{\mathbf{X}=\mathbf{X_{0}}}=a_{0}\hat{\boldsymbol\Phi}_0.
\end{equation}
with $\hat{\boldsymbol\Phi}_{0}$ the unit vector tangential to the limit cycle (circle of fixed points) at the fixed point $\mathbf{X}_{0}$. The second eigenvector has a negative eigenvalue, and corresponds to the relaxation onto the limit cycle. It is most easily derived from the equations in polar (i.e.\ magnitude-phase) coordinates, Eqs.~(\ref{twoEquations}): the eigenvalue is $f_{a}'(a_{0})$ and the eigenvector is
\begin{equation}
\label{Eqn: e1}
\mathbf{e}_{1}\propto f_{a}'(a_{0})\hat{\mathbf{a}}_0+a_{0}f_{\Phi}'(a_{0})\hat{\boldsymbol\Phi}_0,
\end{equation}
with $\hat{\mathbf{a}}_0$ the radial direction at the fixed point (so that the fixed point is $\mathbf{X}_{0}(\Phi_{0})=a_{0}\hat{\mathbf{a}}_0$).

Now consider the term $\mathbf{J}\cdot\mathbf{X_{1}}$ in Eq.~(\ref{Eqn: stochastic X1}). If we expand $\mathbf{X_{1}}$ in components along $\mathbf e_{0}, \mathbf e_{1}$, the Jacobian kills the component along the zero eigenvalue direction $\mathbf{e}_{0}$ leaving just the component along $\mathbf{e}_{1}$ multiplied by $\lambda_{1}$. Therefore, if we multiply Eq.~(\ref{Eqn: stochastic X1}) on the left by a vector $\mathbf{e}_{0}^{\dagger}$ that is perpendicular to $\mathbf e_{1}$ so that $\mathbf{e}_{0}^{\dagger}\cdot\mathbf e_{1}=0$, the equation becomes
\begin{equation}
\label{Eqn: X1 secular}
\frac{d}{dT}(\mathbf{e}_{0}^{\dagger}\cdot\mathbf{X_{1}})=-(\mathbf{e}_{0}^{\dagger}\cdot\mathbf{e}_{0})\frac{d\Phi}{dT}+\mathbf{e}_{0}^{\dagger}\cdot\boldsymbol\Xi.
\end{equation}
The component $\mathbf{e}_{0}^{\dagger}\cdot\mathbf{X_{1}}$ of $\mathbf X_{1}$ will grow to large values over long times, violating the assumption that $\mathbf X_{1}$ is a small correction, unless the right hand side of Eq.~(\ref{Eqn: X1 secular}) is zero. This secular condition gives the phase evolution equation. If we choose the normalization of $\mathbf e_{0}^{\dagger}$ to be $\mathbf e_{0}^{\dagger}\cdot\mathbf e_{0}=1$, then $\mathbf e_{0}^{\dagger}$  defines the \emph{phase sensitivity vector} $\mathbf V_{\perp}$, and the stochastic phase evolution equation is
\begin{equation}
\label{Eqn: secular phase}
\frac{d\Phi}{dT}=\mathbf{V}_{\perp}\cdot\boldsymbol\Xi.
\end{equation}
This result corresponds to the intuitive understanding that, since deviations along $\mathbf{e}_{1}$ relax back to the fixed point and do not change the phase $\Phi$, it is only the component of the noise perpendicular to this direction, i.e. along $\mathbf e_{0}^{\dagger}\equiv\mathbf V_{\perp}$, that contribute to the phase evolution. The precise way in which this happens is specified by Eq.~(\ref{Eqn: secular phase}). Using Eqs.~(\ref{Eqn: e0},\ref{Eqn: e1}), the explicit result for the phase sensitivity vector is
\begin{equation}
\mathbf{V}_{\perp}=-\frac{f_{\Phi}'(a_{0})}{f_{a}'(a_{0})}\hat{\mathbf{a}}_0+\frac{1}{a_{0}}\hat{\boldsymbol\Phi}_0.
\end{equation}
Note that $\mathbf V_{\perp}$ is \emph{not} along the limit cycle in general. To return to the magnitude-phase components of \S\ref{sec: oscillator phase noise}, the component of $\mathbf{V}_{\perp}$ along $\hat{\boldsymbol\Phi}_0$ is multiplied by $a_{0}$: this gives Eq.~(\ref{zeroMode}) in the main text.

The theorems of linear algebra tell us that $\mathbf e_{0}^{\dagger}$ is the zero-eigenvalue adjoint eigenvector (hence the notation), and so can be obtained as the eigenvector of the adjoint Jacobian: this provides a useful way for obtaining the vector in higher dimensional situations, but is not necessary here.

\section{Derivation of the slow dynamics noise spectrum}
\label{Appendix: slow noise}

The Fourier transform of the autocorrelation Eq.~(\ref{Xi_iXi_j}) with Eqs.~(\ref{Xi FT},\ref{cartezian noise}) is
\begin{eqnarray}\label{}
    &&\langle\tilde{\Xi}_{R}(\Omega)\tilde{\Xi}_{R}(\Omega')\rangle=\int_{-\infty}^{\infty}dT\int_{-\infty}^{\infty}dT'\langle\Xi_{R}(T)\Xi_{R}(T')\rangle\nonumber\\ &\times&e^{-i\Omega T}e^{-i\Omega'T'}\nonumber\\
    &=&\frac{1}{\pi^{2}}\int_{-\infty}^{\infty}dT\int_{-\infty}^{\infty}dT'\int_{\varepsilon^{-1}T- \pi}^{\varepsilon^{-1}T+ \pi}dt\int_{\varepsilon^{-1}T'- \pi}^{\varepsilon^{-1}T'+ \pi}dt'\nonumber\\
    &\times&\langle\xi(t)\xi(t')\rangle\cos(t+\psi_{N})\cos(t'+\psi_{N})e^{-i\Omega T}e^{-i\Omega'T'}\nonumber\\
    &=&\frac{\varepsilon^{2}}{ \pi^{2}}\int_{-\infty}^{\infty}dt''\int_{-\infty}^{\infty}dt'''\int_{t''- \pi}^{t''+ \pi}dt\int_{t'''- \pi}^{t'''+ \pi}dt'\\
    &\times&\langle\xi(t)\xi(t')\rangle\cos(t+\psi_{N})_{N}\cos(t'+\psi_{N})e^{-i\Omega \varepsilon t''}e^{-i\Omega'\varepsilon t'''}.\nonumber
\end{eqnarray}
Changing the order of the $t''$ and the $t$ integration gives
\begin{eqnarray}\label{}
    &&\int_{-\infty}^{\infty}dt''\int_{t''- \pi}^{t''+ \pi}dt e^{-i\Omega \varepsilon t''}
    \rightarrow\int_{-\infty}^{\infty}dt\int_{t- \pi}^{t+ \pi}dt''e^{-i\Omega \varepsilon t''}\nonumber\\
    &=&\frac{2}{\Omega\varepsilon}\sin\left(\pi\varepsilon\Omega\right)\int_{-\infty}^{\infty}dte^{-i\Omega \varepsilon t},
\end{eqnarray}
and doing the same with $t'''$ and $t'$ gets us to
\begin{eqnarray}\label{realcorr}
    &&\langle\tilde{\Xi}_{R}(\Omega)\tilde{\Xi}_{R}(\Omega')\rangle=\frac{4}{\pi^{2}\Omega\Omega' }\sin\left(\pi\varepsilon\Omega\right)\sin\left(\pi\varepsilon\Omega'\right)\nonumber\\
    &\times&\int_{-\infty}^{\infty}dt\int_{-\infty}^{\infty}dt'\langle\xi(t)\xi(t')\rangle\cos(t+\psi_{N})\cos(t'+\psi_{N})\nonumber\\
    &\times&e^{-i\Omega \varepsilon t}e^{-i\Omega'\varepsilon t'},
\end{eqnarray}
and similarly
\begin{eqnarray}\label{imagcorr}
    &&\langle\tilde{\Xi}_{I}(\Omega)\tilde{\Xi}_{I}(\Omega')\rangle=\frac{4}{\pi^{2}\Omega\Omega' }\sin\left(\pi\varepsilon\Omega\right)\sin\left(\pi\varepsilon\Omega'\right)\nonumber\\
    &\times&\int_{-\infty}^{\infty}dt\int_{-\infty}^{\infty}dt'\langle\xi(t)\xi(t')\rangle\sin(t+\psi_{N})\sin(t'+\psi_{N})\nonumber\\
    &\times&e^{-i\Omega \varepsilon t}e^{-i\Omega'\varepsilon t'},
\end{eqnarray}
and
\begin{eqnarray}\label{crosscorr}
    &&\langle\tilde{\Xi}_{R}(\Omega)\tilde{\Xi}_{I}(\Omega')\rangle=-\frac{4}{\pi^{2}\Omega\Omega' }\sin\left(\pi\varepsilon\Omega\right)\sin\left(\pi\varepsilon\Omega'\right)\nonumber\\
    &\times&\int_{-\infty}^{\infty}dt\int_{-\infty}^{\infty}dt'\langle\xi(t)\xi(t')\rangle\cos(t+\psi_{N})\sin(t'+\psi_{N})\nonumber\\
    &\times&e^{-i\Omega \varepsilon t}e^{-i\Omega'\varepsilon t'}.
\end{eqnarray}
Putting in the cyclostationary noise Eq.~(\ref{noise correlation}) gives
\begin{eqnarray}\label{}
    &&\langle\tilde{\Xi}_{R}(\Omega)\tilde{\Xi}_{R}(\Omega')\rangle=\frac{4}{\pi^{2}\Omega\Omega' }\sin\left(\pi\varepsilon\Omega\right)\sin\left(\pi\varepsilon\Omega'\right)\nonumber\\
    &\times&\sum_l\bigg[\int_{-\infty}^{\infty}dsR_{l}(s)\cos se^{i\Omega'\varepsilon s}\nonumber\\
    &\times&\int_{-\infty}^{\infty}dt\cos^2(t+\psi_{N})e^{-i[(\Omega+\Omega')\varepsilon-l] t}\nonumber\\
    &+&\int_{-\infty}^{\infty}dsR_{l}(s)\sin se^{i\Omega'\varepsilon}\\
    &\times&\int_{-\infty}^{\infty}dt\cos(t+\psi_{N})\sin(t+\psi_{N})e^{-i[(\Omega+\Omega')\varepsilon-l] t}\bigg],\nonumber
\end{eqnarray}
which then allows us to perform the $dt$ integration and get
\begin{eqnarray}\label{}
    &&\langle\tilde{\Xi}_{R}(\Omega)\tilde{\Xi}_{R}(\Omega')\rangle=\frac{2}{\pi\Omega\Omega'}\sin\left(\pi\varepsilon\Omega\right)\sin\left(\pi\varepsilon\Omega'\right)\nonumber\\
    &\times&
    \sum_l\bigg\{\int_{-\infty}^{\infty}dsR_{l}(s)\cos se^{i\Omega'\varepsilon s}\times\nonumber\\
    &&[2\delta(l-\varepsilon(\Omega+\Omega'))+\delta(l+2-\varepsilon(\Omega+\Omega'))e^{2i\psi_{N}}\nonumber\\
    &+&\delta(l-2-\varepsilon(\Omega+\Omega'))e^{-2i\psi_{N}}]\nonumber\\
    &-&i\int_{-\infty}^{\infty}dsR_{l}(s)\sin se^{i\Omega'\varepsilon s}\times\nonumber\\
    &&[\delta(l+2-\varepsilon(\Omega+\Omega'))e^{2i\psi_{N}}\nonumber\\
    &-&\delta(l-2-\varepsilon(\Omega+\Omega'))e^{-2i\psi_{N}}]\bigg\}.
\end{eqnarray}
In the small $\varepsilon$ limit, we can make the approximation $\delta(l-m -\epsilon(\Omega+\Omega'))\to\delta_{l,m}\delta(\Omega+\Omega')/\epsilon$ which gives
\begin{equation}
\langle\tilde{\Xi}_{R}(\Omega)\tilde{\Xi}_{R}(\Omega')\rangle=2\pi \varepsilon\delta(\Omega+\Omega')S_{RR}(\Omega),
\end{equation}
with
\begin{eqnarray}
    S_{RR}(\Omega)&=&Q_{0}(\varepsilon\Omega-1)+Q_{0}(\varepsilon\Omega+1)\\
    &+&Q_{-2}(\varepsilon\Omega+1)e^{2i\psi_{N}}+Q_{2}(\varepsilon\Omega-1)e^{-2i\psi_{N}}.\nonumber
\end{eqnarray}
where Eq.~(\ref{R-Q}) is used to get the last expression.
Using the symmetry Eqs.~(\ref{Qsymmetry}), and repeating this calculation for the other correlation functions gives
\begin{eqnarray}\label{spec real imaginary and cross}
    S_{RR}(\Omega)&=&Q_{0}(\varepsilon\Omega-1)+Q_{0}(\varepsilon\Omega+1)\\
    &+&2\textmd{Re}[Q_{2}(\varepsilon\Omega-1)e^{-2i\psi_{N}}],\nonumber\\
    S_{II}(\Omega)&=&Q_{0}(\varepsilon\Omega-1)+Q_{0}(\varepsilon\Omega+1)\nonumber\\
    &-&2\textmd{Re}[Q_{2}(\varepsilon\Omega-1)e^{-2i\psi_{N}}],\nonumber\\
    S^s_{RI}(\Omega)&=&S_{RI}(\Omega)+S_{IR}(\Omega)=4\textmd{Im}[Q_{2}(\varepsilon\Omega-1)e^{-2i\psi_{N}}].\nonumber
\end{eqnarray}
The expressions Eq.~(\ref{spectrumHarmonics}) for $Q_{n}$ gives the important result that the slow noise spectral densities $S_{RR}(\Omega)$ etc.\ are \emph{independent} of the phase $\Phi$ of the complex amplitude, so that the slow noise is stationary, independent of the evolution of the phase of the oscillator in the slow time scale.

Typically, since the sustaining part of the closed loop system has a broad frequency response, we might not expect the noise spectra $Q_{n}(\omega)$ to have significant structure on the small frequency scale $\varepsilon\Omega$, so that these terms in the arguments of Eqs.~(\ref{spec real imaginary and cross}) could be neglected, consistent with the neglecting the other terms in $\varepsilon$. This is indeed the case for broadband noise sources, such as white noise. However, the noise itself may induce a nontrivial dependence of $Q_{n}(\omega)$ on small frequency changes. This occurs for amplifying systems producing 1/f noise, which becomes large for small frequencies: the up conversion of the low frequency noise by mixing with the carrier signal via the amplifier nonlinearity leads to a significant dependence of $Q_{\pm 2},Q_{0}$ on the frequency deviation from frequency $\pm 1$, and the $\varepsilon\Omega$ terms in Eqs.~(\ref{spec real imaginary and cross}) cannot be neglected in this case. To make further progress we treat these two cases in turn.

\subsection{Broadband noise}
For broadband noise, we may indeed ignore the $\varepsilon\Omega$ terms in Eqs.~(\ref{spec real imaginary and cross}). Then substituting Eq.~(\ref{spectrumHarmonics}) into these expressions gives
\begin{eqnarray}
\label{specs broadband appendix}
S_{RR}(\Omega)&=&2\big\{\sum_{n}Q_{s_{0}}(n)|H_{n-1}(-n)|^{2}\\
&+&\textmd{Re}[e^{2i\bar\psi_{N}}\sum_{n}Q_{s_{0}}(n)H_{n-1}(-n)H_{n+1}^{*}(-n)]\big\},\nonumber\\
S_{II}(\Omega)&=&2\{\sum_{n}Q_{s_{0}}(n)|H_{n-1}(-n)|^{2}\nonumber\\
&-&\textmd{Re}[e^{2i\bar\psi_{N}}\sum_{n}Q_{s_{0}}(n)H_{n-1}(-n)H_{n+1}^{*}(-n)]\},\nonumber\\
S_{RI}^{s}(\Omega)&=&-4\textmd{Im}[e^{2i\bar\psi_{N}}\sum_{n}Q_{s_{0}}(n)H_{n-1}(-n)H_{n+1}^{*}(-n)],\nonumber
\end{eqnarray}
with $\bar\psi_{N}=\Phi_{N}+\tfrac{\pi}{2}$. These are Eqs.~(\ref{specs broadband}) of the main text.

\subsection{1/f noise}

For 1/f noise sources, the most important terms in the slow noise are given by Eqs.~(\ref{spec real imaginary and cross}) and then restricting $n$ in the sum in Eq.~(\ref{spectrumHarmonics}) so that $n+\omega=O(\varepsilon)$. We can also neglect the dependence of the $H_{n}$ on the small frequency $\varepsilon\Omega$ in these terms so that $H_{n}[\pm(\omega+n)]\simeq H_{n}(0)$. This gives
\begin{eqnarray}
Q_{0}(\varepsilon\Omega\pm 1)&\simeq&Q_{s_{0}}(\varepsilon\Omega)|H_{1}(0)|^{2},\\
Q_{\pm 2}(\varepsilon\Omega\mp1)&\simeq& e^{\pm 2i\Phi}Q_{s_{0}}(\varepsilon\Omega)[H_{\pm 1}(0)]^{2}.\nonumber
\end{eqnarray}
Writing $H_{1}(0)=H_{-1}^{*}(0)=|H_{1}(0)|e^{i\phi_{H}}$ and substituting into Eqs.~(\ref{spec real imaginary and cross}) gives
\begin{eqnarray}
\label{specs 1/f appendix}
S_{RR}(\Omega)&=&2Q_{s_{0}}(\varepsilon\Omega)|H_{1}(0)|^{2}\{1+\cos[2(\phi_{H}-\bar\psi_N)]\},\nonumber\\
S_{II}(\Omega)&=&2Q_{s_{0}}(\varepsilon\Omega)|H_{1}(0)|^{2}\{1-\cos[2(\phi_{H}-\bar\psi_N)]\},\nonumber\\
S_{RI}^{s}(\Omega)&=&4Q_{s_{0}}(\varepsilon\Omega)|H_{1}(0)|^{2}\sin[2(\phi_{H}-\bar\psi_N)].
\end{eqnarray}
These are Eqs.~(\ref{specs 1/f}) of the main text. The neglected $n\ne 0$ terms in Eq.~(\ref{spectrumHarmonics}) will give an additional white contribution to the slow noise corresponding to the $n\ne 0$ terms in Eqs.~(\ref {specs broadband appendix})
\begin{eqnarray}
\label{specs 1/f correction appendix}
S_{RR}(\Omega)&=&2\big\{\sum_{n\ne0}Q_{s_{0}}(n)|H_{n-1}(-n)|^{2}\\
&+&\textmd{Re}[e^{2i\bar\psi_{N}}\sum_{n}Q_{s_{0}}(n)H_{n-1}(-n)H_{n+1}^{*}(-n)]\big\},\nonumber\\
S_{II}(\Omega)&=&2\big\{\sum_{n\ne0}Q_{s_{0}}(n)|H_{n-1}(-n)|^{2}\nonumber\\
&-&\textmd{Re}[e^{2i\bar\psi_{N}}\sum_{n\ne0}Q_{s_{0}}(n)H_{n-1}(-n)H_{n+1}^{*}(-n)]\big\},\nonumber\\
S_{RI}^{s}(\Omega)&=&-4\textmd{Im}[e^{2i\bar\psi_{N}}\sum_{n}Q_{s_{0}}(n)H_{n-1}(-n)H_{n+1}^{*}(-n)].
\nonumber
\end{eqnarray}

As mentioned above, for these 1/f noise calculations we have retained the $O(\varepsilon)$ corrections terms in the frequency arguments in Eq.~(\ref{spec real imaginary and cross}) but not in the prefactors. Since there are also $O(\varepsilon)$ deviations of the oscillator frequency from the linear resonance frequency, it might be a concern that this is not a consistent approximation. In Ref.~\cite{kenig1/f} we calculate the slow noise resulting from 1/f noise sources explicitly retaining these terms, and obtain the same results as in Eqs.~(\ref{specs 1/f appendix}).

\section*{Acknowledgment}
This research was supported by DARPA through the DEFYS program and the National Science Foundation under Grant No. DMR-1003337. The authors thank L.~G.~Villanueva, R.~B.~Karabalin, M.~H.~Matheny, Ron Lifshitz, and M.~L.~Roukes for useful discussions.

\ifCLASSOPTIONcaptionsoff
  \newpage
\fi

\bibliographystyle{IEEEtran}

\bibliography{AmplitudeNoise}

\end{document}